\documentclass[aps,prb,twocolumn,superscriptaddress,reprint]{revtex4-1}
\usepackage{graphicx}
\usepackage[utf8]{inputenc}
\usepackage{amsmath}
\usepackage{siunitx}

\newcommand{\set}[1]{\left\{#1\right\}}
\newcommand{\xLi}{x_\mathrm{Li}}
\newcommand{\hrpa}{{\widetilde{\nu}}}
\newcommand{\angstrom}{\mbox{\normalfont\AA}}

\newcommand{\chem}[1]{\ensuremath{\mathrm{#1}}}

\begin{document}

\title{Lattice-Geometry Effects in Garnet Solid Electrolytes:\\ A Lattice-Gas Monte Carlo Simulation Study}
\author{Benjamin J. Morgan}
\affiliation{Department of Chemistry, University of Bath, Claverton Down, Bath, BA2 7AY}

\date{\today}

\begin{abstract}
Ionic transport in solid electrolytes can often be approximated as ions performing a sequence of hops between distinct lattice sites. If these hops are uncorrelated, quantitative relationships can be derived that connect microscopic hopping rates to macroscopic transport coefficients; i.e.\ tracer diffusion coefficients and ionic conductivities. In real materials, hops are uncorrelated only in the dilute limit. At non-dilute concentrations the relationships between hopping frequency, diffusion coefficient, and ionic conductivity deviate from the random walk case, with this deviation quantified by single-particle and collective correlation factors, $f$ and $f_\mathrm{I}$. These factors vary between materials, and depend on the concentration of mobile particles, the nature of the interactions, and the host lattice geometry. 
Here we study these correlation effects for the garnet lattice using lattice-gas Monte Carlo simulations. We find that for non-interacting particles (volume exclusion only) single-particle correlation effects are more significant than for any previously studied three-dimensional lattice. This is attributed to the presence of two-coordinate lattice sites, which causes correlation effects intermediate between typical three-dimensional and one-dimensional lattices. Including nearest-neighbour repulsion and on-site energies produces more complex single-particle correlations and introduces collective correlations. We predict particularly strong correlation effects at $\xLi=3$ (from site energies) and $\xLi=6$ (from nearest-neighbour repulsion), where $\xLi=9$ corresponds to a fully occupied lithium sublattice. Both effects are consequences of ordering of the mobile particles. Using these simulation data, we consider tuning the mobile ion stoichiometry to maximise the ionic conductivity, and show that the ``optimal'' composition is highly sensitive to the precise nature and strength of the microscopic interactions.
Finally, we discuss the practical implications of these results in the context of lithium garnets and other solid electrolytes.
\end{abstract}

\maketitle

\section{Introduction}

The ability of solid electrolytes to conduct electric charge through ion transport is central to their use in devices such as fuel cells and solid-state lithium-ion batteries.\cite{BachmanEtAl_ChemRev2016, ManthiramEtAl_NatRevMater2017,GoodenoughAndSingh_JElectrochemSoc2015, MalavasiEtAl_ChemSocRev2010} In both cases, solid electrolytes with high ionic conductivities are desirable. In fuel cells high conductivities allow lower operating temperatures, reducing running costs and increasing operating lifetimes. In solid-state batteries high conductivities allow faster charging rates and higher power outputs. Ionic conductivities depend on a number of factors, including the crystal structure, the chemical composition, and the concentration of mobile ions.\cite{VanDerVenEtAl_AccChemRes2013} Developing a quantitative understanding of how these factors interact is key to developing high conductivity solid electrolytes for use in high performance electrochemical devices.

Solid electrolytes can be considered to comprise two distinct sets of ions: ``fixed'' ions that vibrate about their crystallographic sites, and ``mobile'' ions that can move through the system. The fixed ion positions define a network of diffusion pathways through which the mobile ions move. Solid electrolytes with common crystal structures have diffusion networks that are topologically equivalent, while electrolytes with different crystal structures have topologically distinct diffusion paths. While much research into solid electrolytes focusses on understanding differences in ionic conductivities within specific structural families, a complementary question considers how differences in crystal structure, and hence diffusion network topology, affect ionic transport.

Diffusion pathway geometries are defined by crystal structure, and therefore are  a microscopic property of specific materials. The performance of solid electrolytes in devices, however, is characterised by macroscopic transport coefficients: diffusion coefficients and ionic conductivities; which describe ensemble averages over all microscopic diffusion processes. Understanding the differences in ionic conductivity between solid electrolytes depends on resolving the quantitative relationships that link these two perspectives; in doing so, connecting the microscopic picture of specific ion-diffusion mechanisms to the macroscopic properties of long-ranged mass and charge transport.

In many solid electrolytes, the microscopic transport of ions can be approximated as a sequence of discrete ``hops'' between distinct lattice sites.\cite{[{Describing ionic transport as sequences of discrete hops breaks down for ``superionic'' solid electrolytes, with extremely mobile ions. 
The set of criteria for considering ionic transport to operate in a particle hopping regime are discussed by Catlow in }]  [{}] Catlow_SolidStateIonics1983} If these hops are \emph{independent}, every ion follows a random walk. 
The tracer diffusion coefficient, $D^*$, and ionic conductivity, $\sigma$, can then be expressed in terms of the average hop-rate per atom, $\hrpa$,\footnote{The average hop rate per atom is the inverse of the mean residence time, $\hrpa=1/{\widetilde\tau}$. 
The contribution from each atom is a sum over individual hop rates, $\Gamma_i$, and is therefore related to the ``total rate'' of the kMC method via $\hrpa=\left<Q\right>/N$.\cite{Mehrer_DiffusionBook}} 
via \cite{HowardAndLidiard_RepProgPhys1964,Stoneham_IonicSolidsAtHighTemperatures}
\begin{equation}
  D^* = \frac{1}{6}a^2\hrpa;
  \label{eqn:random_walk_diffusion}
\end{equation}
\begin{equation}
  \sigma = \frac{Cq^2}{kT}\frac{1}{6}a^2\hrpa;
  \label{eqn:random_walk_conductivity}
\end{equation}
where $a$ is the characteristic hop distance, $C$ is the mobile ion concentration, and $q$ is the charge of the mobile ions. 
Equations \ref{eqn:random_walk_diffusion} and \ref{eqn:random_walk_conductivity} can be combined to give the Nernst-Einstein relation, which relates $D^*$ and $\sigma$:
\begin{equation}
  \frac{\sigma}{D^*} = \frac{Cq^2}{kT}.
  \label{eqn:random_walk_nernst_einstein}
\end{equation}
These three equations provide quantitative relationships between the hop-rate, $\hrpa$, tracer diffusion coefficient, $D^*$, and ionic conductivity, $\sigma$. 
Their derivation, however, depends on the assumption of independent hops, which holds only in the limit of very low carrier concentrations, or for fully non-interacting mobile ions.\cite{Murch_SolStatIonics1982}

Practical solid electrolytes typically have high carrier concentrations, and interparticle interactions can be significant. 
Under these conditions, individual hopping probabilities depend on the positions of nearby ions, and hops are no longer statistically independent. 
Instead, ion trajectories are correlated, and the system dynamics deviates from random walk behaviour.\cite{BardeenAndHerring_Imperfections1952, CompaanAndHaven_TransFaradaySoc1958, AllnattAndLidiard_AtomicTransportInSolids,HowardAndLidiard_RepProgPhys1964} Correlations between hops made by any single ion modify the relationship between average hop rate per atom, $\hrpa$, and tracer diffusion coefficient, $D^*$, which becomes
\begin{equation}
  D^* = \frac{1}{6}a^2\hrpa f,
  \label{eqn:correlated_diffusion}
\end{equation}
where $f$ is a single-particle correlation factor that accounts for the deviations from random walk behaviour. 
Correlations between hops made by \emph{different} ions modify the relationship between $\hrpa$ and $\sigma$, which becomes
\begin{equation}
  \sigma = \frac{Cq^2}{kT}\frac{1}{6}a^2\hrpa f_\mathrm{I},
  \label{eqn:correlated_conductivity}
\end{equation}
where $f_\mathrm{I}$ is a collective or ``physical'' correlation factor.\cite{Mehrer_DiffusionBook, Murch_SolStatIonics1982,SatoAndKikuchi_JChemPhys1971} 
The relationship between $\sigma$ and $D^*$ now differs from Nernst-Einstein behaviour (Eqn.~\ref{eqn:random_walk_nernst_einstein}) by the ratio of these correlation factors:
\begin{equation}
  \frac{\sigma}{D^*} = \frac{Cq^2}{kT}\frac{f_\mathrm{I}}{f}.
  \label{eqn:correlated_nernst_einstein}
\end{equation}
The inverse ratio $\frac{f}{f_\mathrm{I}}$ is commonly referred to as the Haven ratio, $H_\mathrm{R}$\cite{Murch_SolStatIonics1982,Akbar_JApplPhys1994}.

Empirical relationships between microscopic hopping rates and macroscopic transport coefficients can be obtained, in principle, by combining experimental data for $\hrpa$, $D^*$, and $\sigma$. 
Ion hopping rates can be measured in NMR or muon spin-relaxation experiments,\cite{WilkeningEtAl_PhysRevLett2006, RuprechtEtAl_PhysChemChemPhys2012, Enciso-MaldonadoEtAl_ChemMater2015,Santibanez-MendietaEtAl_ChemMater2016, NozakiEtAl_SolStatIonics2014,AmoresEtAl_JMaterChemA2016} diffusion coefficients obtained from tracer diffusion experiments,\cite{BaylissEtAl_AdvEnergyMater2014} and ionic conductivities extracted from impedance spectroscopy.\cite{ZeierEtAl_ACSApplMaterInt2014,Lopez-BermudezEtAl_2016} Computational methods provide an increasingly useful complement to experimental studies of solid electrolytes. 
First principles calculations of vibrational partition functions and barrier heights along diffusion pathways can be used to obtain hopping rates \emph{ab initio}.\cite{VanDerVenEtAl_PhysRevB2001, MantinaEtAl_PhysRevLett2008} Molecular dynamics simulations can be used to directly calculate diffusion coefficients and ionic conductivities.\cite{MorganAndMadden_JPhysCondensMat2012} Often, however, one or more of $\set{\hrpa,D^*,\sigma}$ are unknown, and it is necessary to derive these from the other, known, properties. In principle, quantitative conversions between $\set{\hrpa,D^*,\sigma}$ are possible via Eqns.~\ref{eqn:correlated_diffusion}--\ref{eqn:correlated_nernst_einstein}, providing the correlation factors $f$, $f_\mathrm{I}$ (and hence also $H_\mathrm{R}$) are known.

For many simple crystal lattices, the correlation parameters $\set{f, f_\mathrm{I}, H_\mathrm{R}}$ have been calculated.\cite{Friauf_JApplPhys1962,Murch_SolStatIonics1982} For more complex crystal structures, however, these parameters are often still unknown. A common approximation, therefore, is to assume correlation effects can be neglected, which allows the simpler Eqns.~\ref{eqn:random_walk_diffusion}--\ref{eqn:random_walk_nernst_einstein} to be used. This approximation is equivalent to assuming dilute-limit non-interacting behaviour. In solid electrolytes, where ionic motion exhibits strong correlation effects, however, this can introduce quantitative errors when processing data.

\begin{figure}[tb]
  \centering
  \resizebox{8cm}{!}{\includegraphics*{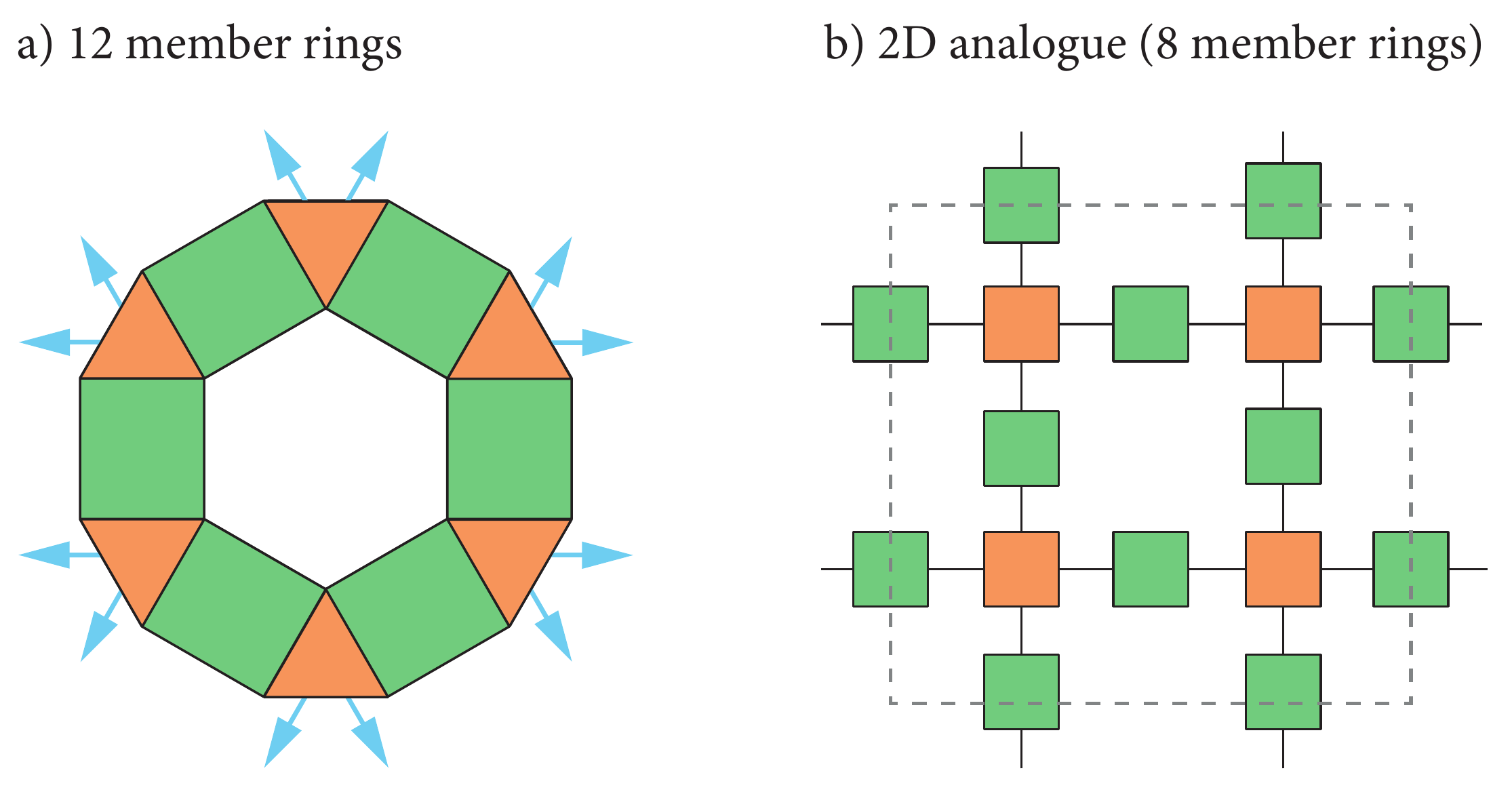}} %
    \caption{\label{fig:garnet_network_schematic}Schematic of the ring structures that constitute the garnet lithium-diffusion network. a) Each ring consists of twelve alternating tetrahedra (orange) and octahedra (blue). 
Arrows show connections to neighbouring rings.\cite{AwakaEtAl_ChemLett2011} b) A two-dimensional analogue, with interconnected eight-membered rings of alternating ``tetrahedra'' and ``octahedra''.}
\end{figure}

In this study, we report lattice-gas Monte Carlo simulation of ionic transport on the garnet lattice, performed to quantify correlation effects for this lattice geometry. The garnet lattice provides a model for diffusion pathways in the ``lithium-garnets'', $\mathrm{Li}_\mathrm{x}M_3M^\prime_2\mathrm{O}_{12}$.\cite{ThangaduraiEtAl_JAmCeramSoc2003, ThangaduraiEtAl_JPhysChemLett2015} This family of solid lithium-ion electrolytes has attracted significant attention as candidate electrolytes for all--solid-state lithium-ion batteries.\cite{BachmanEtAl_ChemRev2016, InadaEtAl_FrontEnergyRes2016,HanEtAl_NatMater2016, RamakumarEtAl_ProgMaterSci2017} The garnet crystal structure has an unusual three-dimensional network of lithium diffusion pathways, consisting of interlocking rings.\cite{AwakaEtAl_ChemLett2011} Each ring comprises twelve alternating tetrahedral and octahedral sites. Each tetrahedral site is coordinated to four octahedral sites, and each octahedral site is coordinated to two tetrahedral sites, with the tetrahedral sites acting as nodal points connecting adjacent rings (Fig.~\ref{fig:garnet_network_schematic}).
Aliovalent substitution of the $M$ and $M^\prime$ cations allows the lithium stoichiometry to be tuned across a broad range. 
A theoretical lithium stoichiometry of $\xLi=9$ corresponds to a fully occupied lithium-site lattice. 
Ionic conductivities vary enormously as a function of $\xLi$, with $\sigma$ increasing by $\sim\!10^9$ between \chem{Li_3Tb_3Te_2O_{12}} and \chem{Li_{6.55}La_3Zr_2Ga_{0.15}O_{12}}.\cite{ThangaduraiEtAl_JPhysChemLett2015, BachmanEtAl_ChemRev2016} It remains an open question how the lithium diffusion coefficient and ionic conductivity vary with lithium stoichiometry. It is also not know to what extent the unusual diffusion pathway topology affects ionic transport. Resolving these questions is critical for the optimisation of ionic conductivity for this family of materials.

Structural considerations and published data both suggest lithium-garnets  exhibit significant correlation effects. The low connectivity of the two-coordinate octahedral sites means blocking effects are expected to be considerable.\cite{AwakaEtAl_ChemLett2011} Short distances between neighbouring lattice sites of $\sim\!2.4\,\mathrm{\angstrom}$ suggest strong lithium--lithium repulsion, expected to be particularly significant at high lithium stoichiometries.\cite{OCallaghanAndCussen_ChemComm2007,OCallaghanAndCussen_SolStatSci2008,Cussen_JMaterChem2010,WangEtAl_SolStatIonics2014} The presence of two non-equivalent sets of lattice sites is also a factor. Non-interacting lithium ions would be expected to occupy octahedral and tetrahedral sites in a 2:1 ratio, reflecting the relative site populations. 
Neutron data, however, show that at low lithium content ($\xLi=3$) only tetrahedral sites are occupied,\cite{OCallaghanEtAl_ChemMater2006} while at higher lithium content ($\xLi=5\to7$) octahedral sites become preferentially occupied.\cite{Cussen_JMaterChem2010,ThangaduraiEtAl_JPhysChemLett2015} Experimental conductivities depend non-linearly on $\xLi$,\cite{ThompsonEtAl_AdvEnergyMater2015} and deviate from ideal values predicted from muon-spin--spectroscopy hopping rates (via Eqn.~\ref{eqn:random_walk_conductivity}).\cite{NozakiEtAl_SolStatIonics2014} Further evidence for correlated transport in lithium garnets comes from computational studies. A variety of correlated diffusion processes have been observed in molecular dynamics simulations,\cite{JalemEtAl_ChemMater2013, MeierEtAl_JPhysChemC2014,KlenkAndLai_PhysChemChemPhys2015, BurbanoEtAl_PhysRevLett2016} and calculated diffusion coefficients and ionic conductivities show non--Nernst-Einstein behaviour ($H_\mathrm{R}<1$).\cite{KlenkAndLai_SolStatIonics2016} These results collectively indicate the existence of significant interactions, either between lithium ions or between these ions and the host lattice. The quantitative effects of correlation in lithium garnets, however, are not known, and consequently studies often assume uncorrelated behaviour when extrapolating between hop rates, diffusion coefficients, and ionic conductivities.\cite{KuhnEtAl_PhysRevB2011,KuhnEtAl_JPhys-CondensMat2011,MiaraEtAl_ChemMater2013,Rustad_arXiv2016,GuEtAl_SolStatIonics2015,NozakiEtAl_SolStatIonics2014,ZeierEtAl_ACSApplMaterInt2014,JalemEtAl_ChemMater2013,AdamsAndRao_JMaterChem2012,DuvelEtAl_JPhysChemC2012,NarayananEtAl_RSCAdv2012,RamzyAndThangadurai_ACSApplMaterInt2010,AmoresEtAl_JMaterChemA2016}

Here we present a computational study of these correlation effects, using lattice-gas kinetic Monte Carlo simulations of diffusion on a garnet lattice, across a range of model Hamiltonians. 
We calculate $f$ and $f_\mathrm{I}$ as functions of lithium stoichiometry, first for a non-interacting volume-exclusion model,\footnote{Here we follow the convention where ``non-interacting'' does not preclude volume-exclusion, where two mobile particles are forbidden from simultaneously occupying a single lattice site.\cite{Kutner_PhysLett1981} This definition is equivalent to all allowed configurations of particles having equal energies.} and then for models that include on-site single-particle energies and/or repulsive nearest-neighbour interactions. 
In addition to self- and collective-correlation factors, we present site occupation populations, diffusion coefficients, and reduced ionic conductivities for this range of simulation models. Our results illustrate how different interactions contribute to non-ideal behaviour, and modify the relationships between particle hopping rate, diffusion coefficient, and ionic conductivity. 

We find that for non-interacting particles (volume exclusion only) single-particle correlation effects are more significant than for any previously studied three-dimensional lattice. This is attributed to the presence of two-coordinate octahedral sites, which produce correlation effects intermediate between typical three- and one-dimensional lattices. Including nearest-neighbour repulsion or on-site energy differences gives more complex single-particle correlation behaviour and introduces collective correlations. In particular, we find strong correlation effects at $\xLi=3$ (due to site energy differences) and $\xLi=6$ (due to nearest-neighbour repulsion). Both effects correspond to mobile particles ordering over the lattice, with associated sharp decreases in diffusion coefficients and ionic conductivities. By analysing our simulation data, we consider the question of tuning the mobile ion stoichiometry to maximise the ionic conductivity. We show this does not have a straightforward answer, and the optimal stoichiometry is highly sensitive to the choice of interaction parameters. Finally, we discuss the practical implications of these results in the context of garnet-structured and other solid electrolytes.

\section{Methods}

\begin{figure*}[tb]
  \centering
  \resizebox{18.1cm}{!}{\includegraphics*{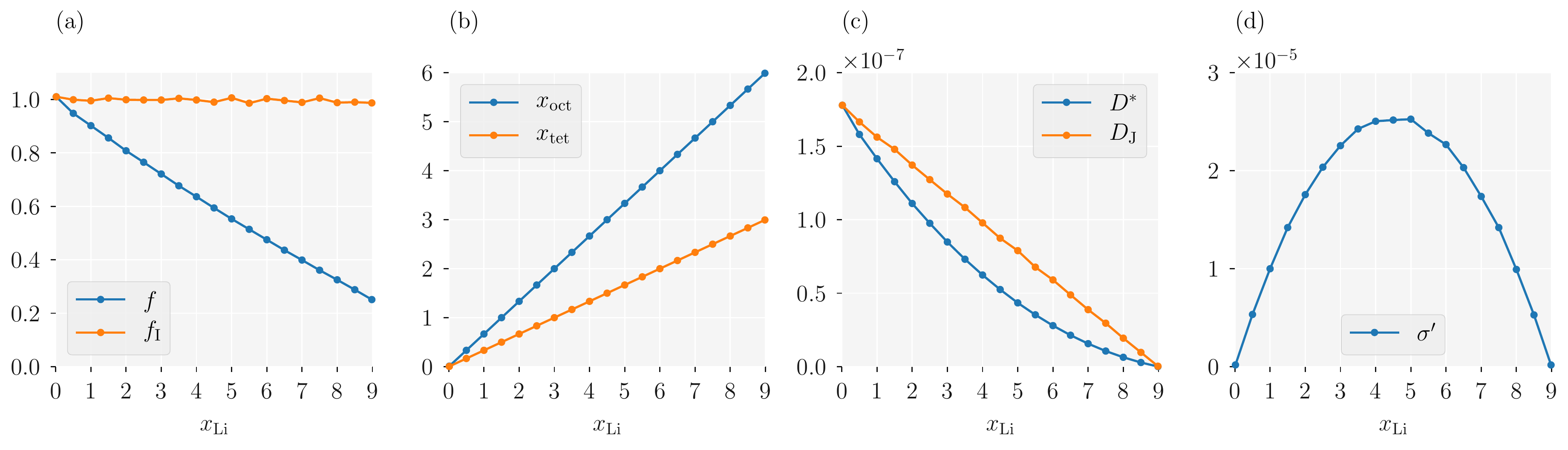}} %
    \caption{\label{fig:non-interacting_data}Non-interacting particles on a garnet lattice: (a) The single-particle correlation factor, $f$, and collective correlation factor, $f_\mathrm{I}$; (b) Average octahedral and tetrahedral site occupations per formula unit, $x_\mathrm{oct}$ and $x_\mathrm{tet}$ (c) Tracer diffusion coefficient, $D^*$, and ``jump'' diffusion coefficient $D_\mathrm{J}$. (d) Reduced ionic conductivity, $\sigma^\prime$.}
\end{figure*}

Lattice-gas Monte Carlo simulations describe the diffusion of a set of mobile ions populating a host lattice, expressed as a graph of interconnected sites.\cite{Trudeau_GraphTheoryBook} 
Every lattice site is either occupied or vacant, and during a simulation the mobile ions hop from site to site. 
These hops are randomly selected, with relative probabilities that satisfy the principle of detailed balance and represent the underlying model Hamiltonian. 
The simplest model considered here is a non-interacting volume-exclusion model.\cite{Kutner_PhysLett1981} Double occupancy of sites is forbidden, and allowed hops are all equally likely. 
The non-interacting model allows the pure geometric effect of the lattice to be evaluated, but neglects other interactions that may be important in experimental systems. 
We therefore also consider the effects of nearest-neighbour interactions between mobile ions, described by a nearest-neighbour repulsion energy, $E_\mathrm{nn}$, and of interactions between single ions and the lattice, described by site-occupation energy differences between tetrahedral and octahedral sites, $E_\mathrm{tet}$, $E_\mathrm{oct}$. 
The energy of any configuration of occupied sites, $\set{j}$, is given by
\begin{equation}
  E = \sum_j n_j^\mathrm{nn}E_\mathrm{nn} + E_\mathrm{site}^j,
\end{equation}
where $n^\mathrm{nn}_j$ is the number of occupied nearest neighbour sites for (occupied) site $j$. 
For interacting systems, the relative probability of hop $i$ depends on the change in total energy if this hop was selected, $\Delta E_i$, according to the scheme of Metropolis \emph{et al.}\cite{MetropolisEtAl_JChemPhys1953};
\begin{equation}
  P_i \propto 
  \begin{cases}
    \mathrm{exp}\left(\frac{-\Delta E_i}{kT}\right),& \mathrm{if}\,\,\Delta E_i > 0 \\
    1,                                        & \mathrm{otherwise.}
  \end{cases}
  \label{eqn:metropolis}
\end{equation}
For our interacting systems, the change in energy for each candidate hop can depend on the change in number of nearest-neighbour interactions and on the change in site occupation energy when moving from a tetrahedral to octahedral site (or vice versa):
\begin{equation}
  \Delta E_i = \Delta n_\mathrm{nn}E_\mathrm{nn} \pm \Delta E_\mathrm{site},
\end{equation}
where $\Delta E_\mathrm{site}=E_\mathrm{oct}-E_\mathrm{tet}$.
At each simulation step, one hop is randomly selected according to the set of relative probabilities. 
The corresponding ion is moved, and a new set of relative hop probabilities is generated for the following simulation step.

In the limit of a large number of hops, the tracer- and collective-correlation factors can be evaluated as
\begin{equation}
  f = \frac{\sum_i\left<R^2\right>}{Na^2},
  \label{eqn:tracer_correlation_factor}
\end{equation}
where $\left<R^2\right>$ is the mean-squared displacement of the mobile ions, and $N$ is the total number of hops during the simulation,\cite{VanDerVenEtAl_PhysRevB2001} and
\begin{equation}
  f_\mathrm{I} = \frac{\left|\sum_i R_i\right|^2}{Na^2},
  \label{eqn:collective_correlation_factor}
\end{equation}
where $\sum_i R_i$ is the \emph{net} displacement of all mobile particles. 
In both cases the denominators correspond to the limiting behaviour for uncorrelated diffusion.

To allow time-dependent properties to be evaluated, such as average site occupations and transport coefficients, we perform our simulations within a rejection-free kinetic Monte Carlo scheme.\cite{Voter_RadiationEffectsInSolids2007} 
At each simulation step, $k$, the set of relative hop probabilities, $\set{P_{i,k}}$, are converted to rates, $\set{\Gamma_{i,k}}$, by scaling by a common prefactor $\nu^\prime=10^{13}\,\mathrm{s}^{-1}$. 
After selecting a hop, the simulation time is updated by $\Delta t = Q_k^{-1}\ln\left(1/u\right)$, where $Q_k$ is the ``total rate''; $Q_k=\sum_i \Gamma_{i,k}$, and $u$ is a uniform random number $u\in\left(0,1\right]$. 

Our lattice-gas kinetic Monte Carlo simulations were performed using the \texttt{lattice\_mc} code.\cite{Morgan_JOSS2017} 
Simulations were performed for an ideal cubic $2\times2\times2$ garnet lattice, with 384 octahedral sites and 192 tetrahedral sites. 
The lattice-site coordinates were generated from the cubic high-temperature \chem{Li_7La_3Zr_2O_{12}} (LLZO) structure,\footnote{ICSD \#422259.\cite{AwakaEtAl_ChemLett2011}} using the centres of the octahedra and tetrahedra defined by the oxide sublattice. 
In cubic LLZO, each octahedron available to lithium contains a ``split'' pair of distorted $96h$ sites, separated by $0.81\,\mathrm{\angstrom}$. 
The construction used here considers each octahedron as a single ideal $48g$ site. 
The graph of diffusion pathways includes connections between nearest-neighbour sites only, i.e.\ all connections are between neighbouring tetrahedra--octahedra pairs. For each simulation, $n_\mathrm{Li}$ mobile ions are initially randomly distributed across the lattice sites. We perform 1,000 simulation steps for equilibration, followed by 10,000 production steps. 

For each set of model parameters, $\set{E_\mathrm{nn}, \Delta E_\mathrm{site}}$, simulations were performed across the full range of possible lithium stoichiometry. For a $2\times2\times2$ garnet supercell, the maximum lithium content of $\xLi=9$ corresponds to $n_\mathrm{Li}=576$. 
For each set of interaction parameters, data were collected as an average over 5,000 independent trajectories. 

\section{Results}
\subsection{Non-Interacting Particles and Geometric Effects}

We first examine the geometric effect of the garnet lattice by considering non-interacting particles, where any deviations from random walk behaviour are solely due to blocking effects.
Fig.~\ref{fig:non-interacting_data}  shows, as a function of $\xLi$, (a) the calculated self- and collective-correlation factors, $f$ and $f_\mathrm{I}$, (b) average tetrahedral and octahedral site occupations, $n_\mathrm{tet}$ and $n_\mathrm{oct}$, (c) tracer and ``jump'' diffusion coefficients, $D^*$ and $D_\mathrm{J}$, (d) and a reduced ionic conductivity, $\sigma^\prime$ (Eqn.~\ref{eqn:reduced_sigma}).

In the single particle limit, $\xLi\to0$, both correlation factors equal 1. There are no blocking effects, and particles follow a random walk. 
With increasing concentration, however, the single particle diffusion deviates from random walk behaviour. 
The tracer correlation factor, $f$, decreases from $f=1$ in the single particle limit to $f=0.25$ in the single vacancy limit $\xLi\to9$, showing approximately linear dependence on $\xLi$.\footnote{A linear least-squares fit to these data gives $R^2=0.9974$.}

The magnitude of the tracer correlation effect for different lattice geometries can be quantified by considering $f$ in the limit of a single vacancy, $f_\mathrm{v}$. 
Table \ref{tab:vacancy_correlation_factors} presents $f_\mathrm{v}$ values previously calculated for simple three-dimensional lattices,\cite{CompaanAndHaven_TransFaradaySoc1956} and for a one-dimensional chain,\cite{Mehrer_DiffusionBook} alongside our result for the garnet lattice. 
The garnet lattice value of $f_v=0.25$ is less than for all previously studied three-dimensional lattices, and is a factor of two smaller than the next lowest (the diamond lattice). This indicates particularly strong site-blocking effects.  
For a general set of three-dimensional lattices, as the number of nearest neighbours of each lattice site, $z$, decreases, $f_v$ also decreases, as site-blocking effects become more significant.\cite{Friauf_JApplPhys1962}
The garnet lattice has both four-coordinate (tetrahedral) and two-coordinate (octahedral) sites, and ion hopping follows an alternating tet$\to$oct$\to$tet$\to$oct sequence. 
The calculated value of $f_v=0.25$ is halfway between the values for a four-coordinate three-dimensional diamond lattice ($f_v=0.5$) and for the two-coordinate one-dimensional chain ($f_v=0$).\cite{Mehrer_DiffusionBook} 
This suggests that the low value of $f_v$ for the garnet lattice is a consequence of the low coordination of the lattice sites, in particular the local one-dimensional coordination at the octahedral sites, which act as bottlenecks for long-ranged diffusion. 

\begin{table}[htb]
   \begin{center}
     \begin{tabular}{lrl} \hline
        \multicolumn{1}{c}{Lattice} & \multicolumn{1}{c}{$z$} & \multicolumn{1}{c}{$f_v$} \\ \hline
        Face centred cubic\cite{CompaanAndHaven_TransFaradaySoc1956} & 12 & 0.78146 \\
        Body centred cubic\cite{CompaanAndHaven_TransFaradaySoc1956} & 8 & 0.72722 \\
        Simple cubic\cite{CompaanAndHaven_TransFaradaySoc1956} & 6 & 0.65311 \\
        Diamond\cite{CompaanAndHaven_TransFaradaySoc1956} & 4 & 0.5 \\
        Garnet [This work] & 4+2 & 0.25 \\ 
        1D chain\cite{Mehrer_DiffusionBook} & 2 & 0.0 \\ \hline
     \end{tabular}
   \caption{\label{tab:vacancy_correlation_factors}Vacancy correlation factors for some common crystal lattices. $z$ is the number of nearest neighbours for each site in the lattice.}
   \end{center}
 \end{table}

For any non-interacting system, the hops made by \emph{different} particles are uncorrelated, and $f_I=1$ for all $\xLi$; hence $H_\mathrm{R}=f$. 
There are also no correlations between site occupations, and the mobile particles are randomly distributed over the available octahedral and tetrahedral sites, with a 2:1 occupation ratio that reflects the underlying lattice geometry.\footnote{In the dilute limit an ion occupying a four-coordinate tetrahedral site has four possible hops that allow it to escape. An ion occupying a two-coordinate octahedral site has only two possible hops. For the non-interacting system, all hops are equally probably, hence the mean residence time for a tetrahedral site is half that of an octahedral site.}

We also calculate three explicit measures of ionic transport in this system.\footnote{Because the lattice-gas model used here considers hops as barrierless, where hopping probabilities only depend on energy differences between initial and final states, the effective transport coefficients calculated here cannot be directly compared to experimental values. For non-interacting systems, introducing fixed barrier heights for tet$\leftrightarrow$oct hops is equivalent to scaling the hopping prefactor $\nu^\prime$, which preserves \emph{relative} differences in the transport coefficients presented here. 
A more realistic model would need to account for the influence of local site occupations on individual hopping barriers, see e.g.\ Refs.~\onlinecite{SingerAndPeschel_ZPhysikBCondensedMatter1980,VanderVenAndCeder_HandbookofMaterialsModelling2010}, and would give quantitative deviations from the trends presented here for diffusion coefficients and correlation factors.} Fig.~\ref{fig:non-interacting_data}(c) shows the tracer diffusion coefficient, $D^*$ (Eqn.~\ref{eqn:correlated_diffusion}) and the ``jump diffusion coefficient'', $D_\mathrm{J}$,\cite{VanDerVenEtAl_AccChemRes2013} calculated as
\begin{equation}
  D_\mathrm{J}=\frac{\left|\sum_iR_i\right|^2}{6Nt}.
\end{equation}
At a fixed temperature, $D_\mathrm{J}$ is proportional to the mobility, and measures the ease with which the mobile particles collectively migrate. 
Both $D^*$ and $D_\mathrm{J}$ decrease monotonically from $\xLi=0$ to $\xLi=9$ ($x=0\to1$), as progressively fewer vacancies are available to accommodate hopping ions. 
For the non-interacting system there are no correlations between hops made by different particles, and the jump diffusion coefficient is proportional to $(1-x)$ (in the garnet lattice, $x=1$ corresponds to a stoichiometry of $\xLi=9$).\cite{Kutner_PhysLett1981,VanDerVenEtAl_AccChemRes2013}
The tracer diffusion coefficient, however, is affected by correlations between hops made by individual particles, and varies as $D^*\propto(1-x)f$. 
\footnote{$D_\mathrm{J}$ is related to the ionic conductivity (via Eqns.~\ref{eqn:correlated_conductivity} and \ref{eqn:collective_correlation_factor}) and also to the chemical diffusion coefficient, ${\widetilde{D}}$, via the thermodynamic factor, $\Theta$, via ${\widetilde{D}}=D_\mathrm{J}\Theta$, where
\begin{equation}
  \Theta = \cfrac{\partial\left(\cfrac{\mu}{kT}\right)}{\partial \ln x}.
\end{equation}}
The ionic conductivity of a system depends on both the charge-carrier concentration, and the ionic mobility, which is proportional to $D_\mathrm{J}$. We quantify the relative effect of carrier concentration on ionic conductivity by
considering a reduced conductivity, $\sigma^\prime$,\footnote{For a system with a single mobile species, the reduced conductivity is equal to the true ionic conductivity if $(VkT)/(q^2)=1$.} given by
\begin{equation}
  \label{eqn:reduced_sigma}
  \sigma^\prime = xD_\mathrm{J}.
\end{equation}
For any non-interacting system, $\sigma^\prime\propto x\left(1-x\right)$, giving a maximum at $x=0.5$, corresponding to $\xLi=4.5$ in the garnet lattice (Fig.~\ref{fig:non-interacting_data}(d)).

\subsection{Interacting Particles}

\begin{figure*}[tb]
  \centering
  \resizebox{14cm}{!}{\includegraphics*{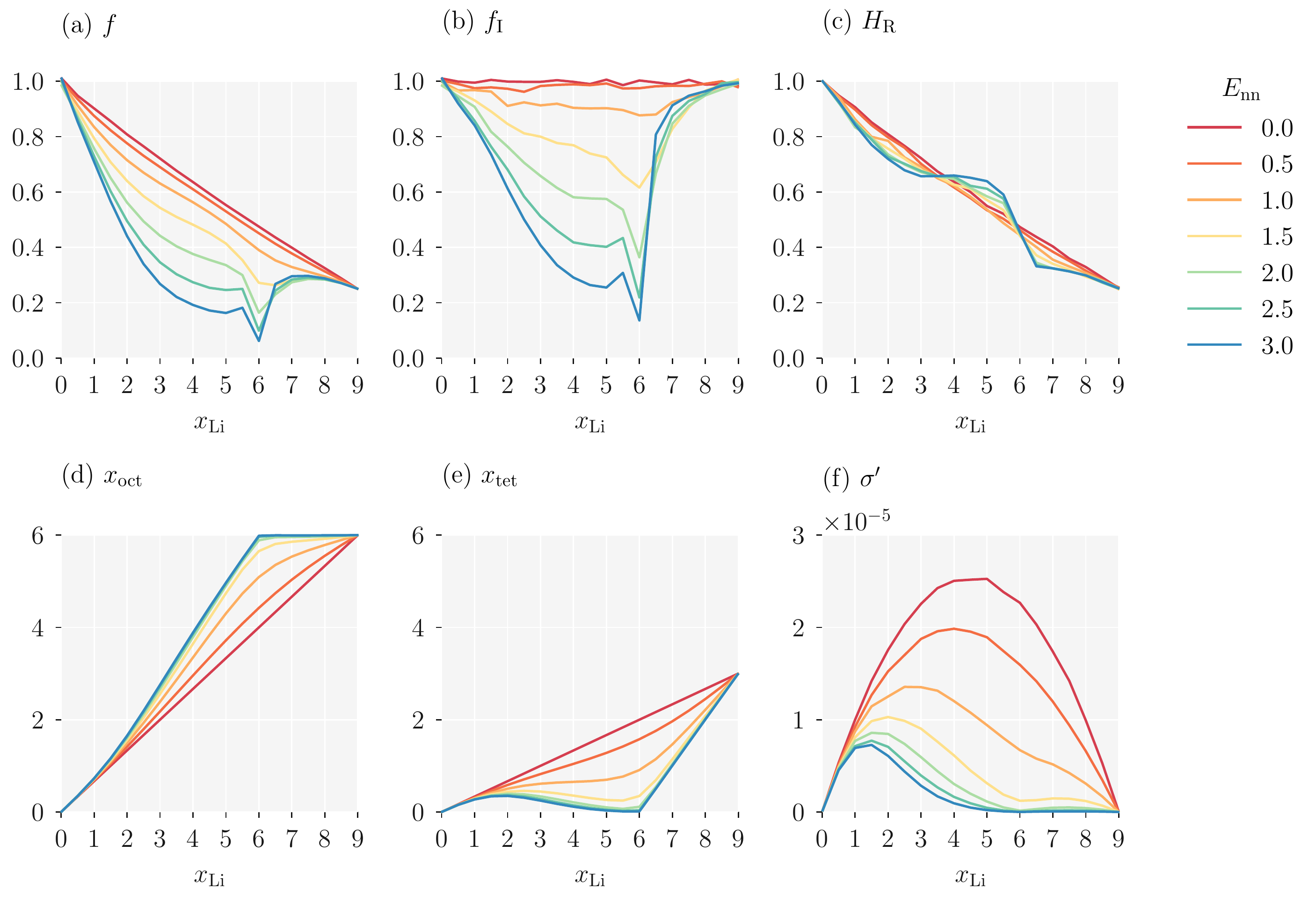}} %
    \caption{\label{fig:nearest_neighbour_data}The effect of nearest-neighbour repulsion between mobile particles on a garnet lattice: (a) single-particle correlation factor, $f$; (b) collective correlation factor, $f_\mathrm{I}$; (c) Haven ratio, $H_\mathrm{R}$; (d) average octahedra occupation, $x_\mathrm{oct}$; (e) average tetrahedra occupation, $x_\mathrm{tet}$; (f) reduced ionic conductivity, $\sigma^\prime$. $E_\mathrm{nn}$ is in multiples of $kT$.}
\end{figure*}

\begin{figure*}[tb]
  \centering
  \resizebox{14cm}{!}{\includegraphics*{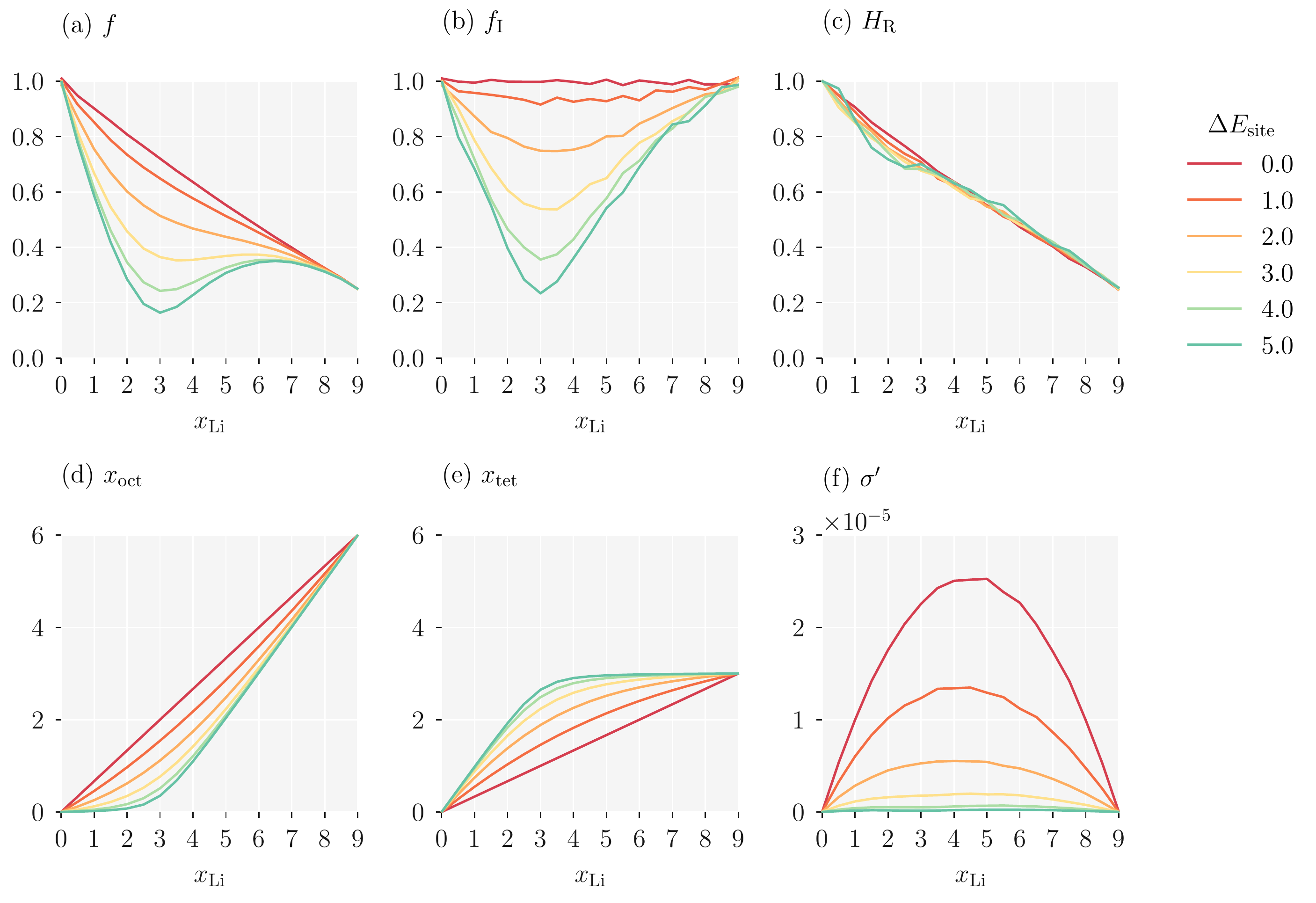}} %
    \caption{\label{fig:site_energies_data}The effect of unequal site occupation energies for mobile particles on a garnet lattice: (a) single-particle correlation factor, $f$; (b) collective correlation factor, $f_\mathrm{I}$; (c) Haven ratio, $H_\mathrm{R}$; (d) average octahedra occupation, $x_\mathrm{oct}$; (e) average tetrahedra occupation, $x_\mathrm{tet}$; (f) reduced ionic conductivity, $\sigma^\prime$. $\Delta E_\mathrm{site}$ is in multiples of $kT$.}
\end{figure*}

The conceptual simplicity of the non-interacting system makes it a useful starting point for understanding the factors affecting ionic transport in different lattices. In particular, purely geometric effects can be resolved.
In real lithium-garnet materials, however, interactions exist between lithium ions, and between lithium ions and the host lattice, and these can significantly affect ion transport. 
Lithium ions are positively charged, and can be expected to experience mutual electrostatic repulsion. The different oxygen-coordination environments of the octahedral and tetrahedral sites can be expected to produce a preference for occupation by lithium at one site versus the other.\cite{WangEtAl_NatMater2015} 
Within the lattice-gas Monte Carlo scheme, we consider these two factors by introducing, first, nearest-neighbour repulsion, and second, an octahedral versus tetrahedral site preference.

\subsubsection{Nearest-neighbour repulsion}

For lithium--lithium repulsion, we consider a simplified model with only nearest-neighbour repulsion. 
The energy of lithium at each site is now proportional to the number of occupied neighbouring sites, and individual hop probabilities depend on whether they increase or decrease the total number of nearest-neighbour pairs. 
Fig.~\ref{fig:nearest_neighbour_data} presents results from simulations performed for $E_\mathrm{nn}=0.0$--$3.0\,kT$. 
Repulsive nearest-neighbour interactions disfavour simultaneous occupation of adjacent pairs of sites, which promotes ordering of particles into alternating  occupied--vacant--occupied--vacant configurations.
This ordering causes the single-particle correlation behaviour to deviate from that of the non-interacting system, and also introduces collective correlations between the mobile ions.\cite{Murch_SolStatIonics1982} $f$ and $f_\mathrm{I}$ both have their non-interacting values in the empty and fully-occupied lattice limits: $x\to0$ and $x\to1$. In a lattice with only one crystallographic site, complete ordering would occur at half--site-occupancy, corresponding to $\xLi=4.5$ for the garnet lattice. $f$ and $f_\mathrm{I}$ approximately follow this trend (Fig.~\ref{fig:nearest_neighbour_data}(a,b)), both decreasing at intermediate $\xLi$ values as $E_\mathrm{nn}$ increases. Superimposed on this general shape, for larger $E_\mathrm{nn}$ values, both correlation factors sharply decrease at $\xLi=6$, i.e.\ two-thirds occupancy. Because $f$ and $f_\mathrm{I}$ do not change uniformly as $E_\mathrm{nn}$ is increased, the Haven ratio $H_\mathrm{R}$ develops structure. Above $\xLi=6$, corresponding to stoichiometries of typical lithium-stuffed garnets, nearest-neighbour repulsion reduces $H_\mathrm{R}$ even further from the already low value for the non-interacting system.

The garnet lattice contains octahedral and tetrahedral sites in a 2:1 ratio. In the non-interacting system, the average site occupancies follow this ratio at all values of $\xLi$ (Fig.~\ref{fig:non-interacting_data}(b)). Introducing repulsive nearest-neighbour interactions increases the probability that octahedra are occupied relative to tetrahedra. Because octahedral sites are two-coordinate, compared to the four-coordinate tetrahedral sites, occupying octahedral sites minimises the number of unfavourable nearest-neighbour interactions. This effect is strongest at two-thirds site occupation ($\xLi=6$) where a sufficiently large $E_\mathrm{nn}$ drives the system into a fully ordered arrangement with all the octahedral sites filled and all the tetrahedral sites empty. 
In this fully ordered system, correlation effects are maximised: a single ion hopping from octahedron to tetrahedron is blocked from further forward motion, and must return to its starting position unless the blocking ion moves first, disrupting the local ordering.\footnote{This second ion will also be blocked unless a third ion moves, and so on, illustrating the requirement that diffusion in this regime becomes highly collective.} Diffusion is only possible for groups of particles undergoing highly correlated collective motion.\cite{MorganAndMadden_PhysRevLett2014,BurbanoEtAl_PhysRevLett2016} Both tracer diffusion and ionic conductivity are strongly reduced compared to their values in the non-interacting system. The collective correlation effects ($f_\mathrm{I}<1$) are visible in the reduced conductivity, $\sigma^\prime$, which decreases relative to the non-interacting system across the full $\xLi$ range, with a particularly strong decrease at $\xLi=6$.

\subsubsection{Asymmetric site-occupation energies}
In the non-interacting model, not only do mobile ions not interact with each other (excepting volume exclusion), but there are no interactions between the mobile ions and the host lattice. Identifying a site as octahedral or tetrahedral only has relevance for defining the connectivity of the lattice graph. Mobile ions show an equal preference for octahedral and tetrahedral sites, with average occupations following a simple 2:1 ratio. This behaviour contrasts with experimental observations. Neutron data for lithium-garnets such as \chem{Li_3Y_3Te_2O_{12}} reveal that at $\xLi=3$ the lithium ions exclusively occupy the tetrahedral sites.\cite{OCallaghanEtAl_ChemMater2006}$^,$\footnote{Several studies of ``lithium-stuffed'' garnets with $\xLi\approx7$ have also reported non-ideal distributions of lithium over tetrahedral and octahedral sites.\cite{XieEtAl_ChemMater2011} Here we focus on the lower concentration $\xLi=3$ data, where additional interactions, such as lithium--lithium repulsion, are expected to play less of a role.} 
This suggests that at relatively low lithium concentrations, there is an energetic penalty for occupying octahedral rather than tetrahedral sites.\footnote{A preference for lithium to occupy tetrahedral rather than octahedral sites mirrors the results of Wang \emph{et al.}, who have shown that for generic fcc and hcp lattices lithium similarly prefers to occupy tetrahedra.\cite{WangEtAl_NatMater2015}}
We model this difference in site-occupation energies by including an on-site term $\Delta E_\mathrm{site} = E_\mathrm{oct} - E_\mathrm{tet}$. To investigate the effect of this ion--lattice interaction on ion dynamics and site occupations we performed a series of simulations for otherwise non-interacting particles, with $\Delta E_\mathrm{site}=0$--$5\,kT$.

The effect of ion--lattice interactions qualitatively mirrors the effect of nearest-neighbour interactions (Fig.~\ref{fig:site_energies_data}). Both single-particle and collective correlation factors are lower then their non-interacting values, average site occupancies deviate from those in the ideal system, and the reduced ionic conductivity decreases. Here, however, the strongest correlations emerge at $\xLi\approx3$. As $\Delta E_\mathrm{site}$ increases, tetrahedral sites are preferentially occupied with respect to octahedral sites, contrasting with the opposite behaviour observed for increasing $E_\mathrm{nn}$. In the limit $T\to0$ this again results in a fully ordered arrangement of ions, now with all the tetrahedral sites filled and all the octahedral sites empty. The Haven ratio, $H_R$, shows less variation compared to the non-interacting result, with only a small decrease for $\xLi<3$.

\subsubsection{Combined site inequality and nearest-neighbour repulsion}

\begin{figure*}[tb]
  \centering
  \resizebox{14cm}{!}{\includegraphics*{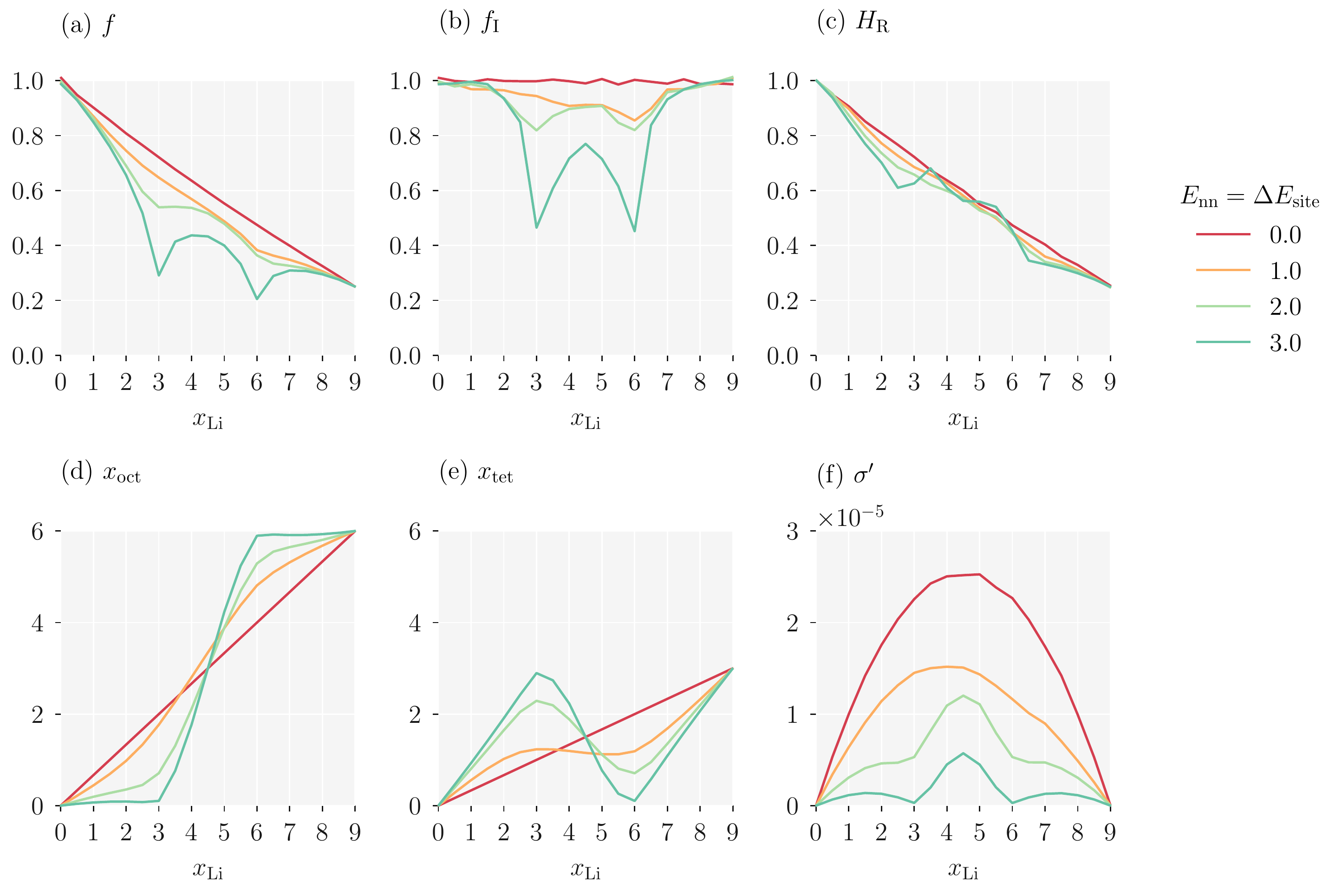}} %
    \caption{\label{fig:both_energies_data}The effect of combined nearest-neighbour repulsion and site-occupation energy differences on a garnet lattice, for $E_\mathrm{nn}=\Delta E_\mathrm{site}$: (a) single-particle correlation factor, $f$; (b) collective correlation factor, $f_\mathrm{I}$; (c) Haven ratio, $H_\mathrm{R}$; (d) average octahedra occupation, $x_\mathrm{oct}$; (e) average tetrahedra occupation, $x_\mathrm{tet}$; (f) reduced ionic conductivity, $\sigma^\prime$. $E_\mathrm{nn}$ and $\Delta E_\mathrm{site}$ are in multiples of $kT$.}
\end{figure*}

In real lithium-garnet electrolytes, lithium ions can be expected to interact with both the host lattice and with each other. To explore the behaviour when both nearest-neighbour and site-occupation interactions are present, we performed simulations to map the $\left\{\xLi, \Delta E_\mathrm{site}, E_\mathrm{nn}\right\}$ parameter space. 
The data from these calculations are presented in Figs.~\ref{fig:correlation_miniplots}--\ref{fig:conductivity_miniplots}. With both interactions present, the ion dynamics and site occupation statistics are  more complex, with specific details that depend on the precise values of both interaction terms. The general features, however, are illustrated by considering the subset $E_\mathrm{nn}=\Delta E_\mathrm{site}$ (Fig.~\ref{fig:both_energies_data}). The correlation factors, $f$ and $f_\mathrm{I}$, both show sharp decreases at $\xLi=3$ and at $\xLi=6$, in both cases corresponding to ordered arrangements of lithium ions throughout the lattice. As in the previous single-interaction models, the ordering at $\xLi=3$ corresponds to filled tetrahedra and empty octahedra (due to $\Delta E_\mathrm{site})$, and the the ordering at $\xLi=6$ corresponds to filled octahedra and empty tetrahedra (due to $E_\mathrm{nn}$). The average site occupation switches sharply from pure tetrahedral occupation to pure octahedral occupation in the range $\xLi=3\to6$. The reduced ionic conductivity, $\sigma^\prime$, is depressed most strongly at lithium stoichiometries corresponding to the ordered arrangements of ions, again, mirroring the results for single interactions. 

\subsubsection{Tuning lithium stoichiometry to maxmimise ionic conductivity}

One challenge regarding lithium garnet solid electrolytes is the question of identifying specific compositions with high ionic conductivities. For garnets with stoichiometries $\mathrm{Li}_xA_3B_2\mathrm{O}_{12}$, the lithium content can be continuously varied by choosing appropriate $A$ and $B$ cations, or by substituting Li$^+$ with small hypervalent cations such as Al$^{3+}$ or Ga$^{3+}$. Different lithium stoichiometries can exhibit very different ionic conductivities. For example, at $\xLi=3$ (e.g.\ Li$_3$Y$_3$Te$_2$O$_{12}$) room temperature conductivities are typically too low to measure\cite{OCallaghanEtAl_ChemMater2006,BachmanEtAl_ChemRev2016,ThangaduraiEtAl_JPhysChemLett2015} 
while for $\xLi\approx6.5$ (e.g.\ Li$_{6.55}$Ga$_{0.15}$La$_3$Zr$_2$O$_{12}$) conductivities as high as $1.3\times10^{-3}\,\mathrm{S}\,\mathrm{cm}^{-1}$ have been reported.\cite{Bernuy-LopezEtAl_ChemMater2014,RettenwanderEtAl_InorgChem2014} One strategy for identifying lithium garnets with high ionic conductivities is to consider whether there is an ``optimal'' lithium stoichiometry that maximises the ionic conductivity.\cite{MuruganEtAl_JElectrochemSoc2008,MuruganEtAl_Ionics2007,RamakumarEtAl_DaltonTrans2015,MiaraEtAl_ChemMater2013,XieEtAl_ChemMater2011,MuruganEtAl_MaterSciEngB2007,OCallaghanAndCussen_ChemComm2007,XuEtAl_PhysRevB2012,ChenEtAl_SciRep2017} A conceptually related question concerns how the ionic conductivity depends on the distribution of lithium ions over tetrahedral and octahedral sites.\cite{ChenEtAl_ChemMater2015,ThangaduraiEtAl_JAmCeramSoc2003,MuruganEtAl_MaterSciEngB2007,OCallaghanAndCussen_ChemComm2007} The lithium distribution is itself a function of the lithium stoichiometry, modulated by the interactions experienced by the lithium ions, as seen above for the model Hamiltonians including nearest-neighbour repulsion and site-occupation energies.
 
For a non-interacting lattice gas, the ionic conductivity varies with the mole fraction of mobile particles, $x$, as
\begin{equation}
  \sigma \propto x \left(1-x\right).
\end{equation}
The $(1-x)$ term is a ``blocking'' factor, due to volume exclusion.\cite{Kutner_PhysLett1981} The conductivity varies parabolically, as seen in the non-interacting system results presented above (Fig.\ \ref{fig:non-interacting_data}(d)). For the lithium garnets this would give a maximum ionic conductivity at $\xLi=4.5$. In real systems, the mobile ions are subject to additional interactions that introduce collective correlations, and the variation in ionic conductivity with mole fraction of mobile particles becomes
\begin{equation}
  \sigma \propto x \left(1-x\right)f_\mathrm{I}.
  \label{eqn:sigma_conc_dependence}
\end{equation}
Because $f_\mathrm{I}$ is itself a function of $x$, this gives non-trivial overall concentration dependence that cannot be described analytically. The concentration dependence of $f_\mathrm{I}$ is an emergent property of the specific interactions the lithium ions are subject to, which indicates that the mobile ion concentration that maximises the ionic conductivity in turn depends on the details of the lithium-ion interactions. 

\begin{figure}[tb]
  \centering
  \resizebox{8cm}{!}{\includegraphics*{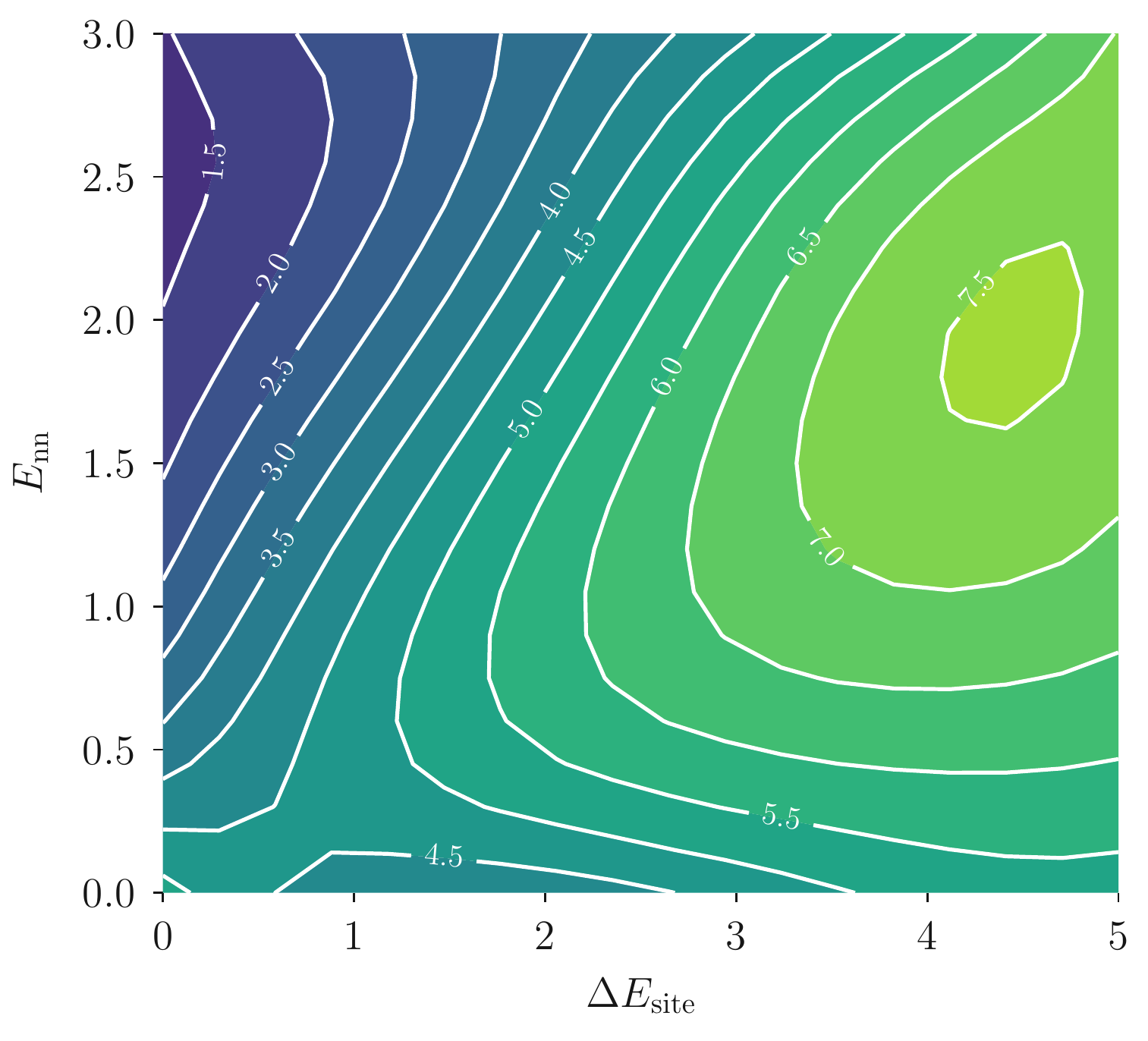}} %
    \caption{\label{fig:max_sigma}Contour plot of the value of $\xLi$ that maximises the reduced ionic conductivity, $\sigma^\prime$, as a function of nearest-neighbour interaction, $E_\mathrm{nn}$, and on-site energy difference, $\Delta E_\mathrm{site}$: $g(E_\mathrm{nn}, \Delta E_\mathrm{site})$, $g = \arg \max \sigma^\prime(\xLi)$.}
\end{figure}

To explore this relationship in the model systems considered here, we can identify the maximum reduced ionic conductivity as a function of lithium stoichiometry; $\arg\max \sigma^\prime(\xLi)$; for each interaction parameter set $\set{E_\mathrm{nn}, \Delta E_\mathrm{site}}$. The resulting surface in parameter space is plotted in Fig.\ \ref{fig:max_sigma}.
As suggested by equation \ref{eqn:sigma_conc_dependence}, the value of $\xLi$ that maximises the ionic conductivity is strongly dependent on the interaction parameter values, due to their effect on the $f_\mathrm{I}(\xLi)$. For the non-interacting system ($E_\mathrm{nn}=0$, $\Delta E_\mathrm{site}=0.0$) $\arg \max \sigma^\prime(\xLi)=4.5$. For non-zero interaction parameters, however, $\arg \max \sigma^\prime(\xLi)$ ranges from $<1.5$ to $>7.5$. Interestingly, the site-occupation energy difference has little effect in the limit of zero nearest-neighbour interactions. As $E_\mathrm{nn}$ increases, $\arg \max \sigma^\prime(\xLi)$ deviates from the non-interacting value. At low values of $\Delta E_\mathrm{site}$, increasing nearest-neighbour repulsion cause the optimal $\xLi$ to decrease. This is associated with the strong suppression of collective ion transport close to $\xLi=6$ (cf.\ Fig.~\ref{fig:nearest_neighbour_data} \& Fig.~\ref{fig:conductivity_miniplots}). At high values of $\Delta E_\mathrm{site}$, however, increasing nearest-neighbour repulsion causes the optimal $\xLi$ to increase, reaching a maximum value of $\sim7.5$. Under these conditions, the preference to occupy tetrahedral sites dominates, and ion transport rates are most strongly decreased close to $\xLi=3$ (cf.\ Fig.~\ref{fig:site_energies_data} \& Fig.~\ref{fig:conductivity_miniplots}).

\section{Summary \& Discussion}

By considering ionic transport in solid electrolytes as effected by particles undergoing random hops, simple analytical relationships can be derived that quantitatively connect microscopic hop rates to macroscopic transport coefficients (cf.\ Eqns.~\ref{eqn:random_walk_diffusion}--\ref{eqn:random_walk_nernst_einstein}). In real solid electrolytes these equations are exact only in the dilute limit. At moderate mobile ion concentrations, ion hops are not independent; instead, they are correlated. The probability of a specific hop occurring depends on the particular arrangement of other nearby ions. Correlations between hops modify the quantitative relationships between hop rates and transport coefficients, with deviations from random walk behaviour expressed via the single-particle and collective correlation factors, $f$ and $f_\mathrm{I}$. Quantifying these correlation factors allows accurate conversions between microscopic (hopping rates) and macroscopic (tracer diffusion coefficients and ionic conductivities) transport data. These factors also provide information about the ionic transport process: the single particle correlation factor quantifies the efficiency with which individual ions move through the electrolyte structure; the collective correlation factor provides an equivalent measure for the efficiency of mass or charge transport.

The simplest cause of correlation effects is volume exclusion, where occupied sites are unavailable to adjacent ions. This effect causes sequential hops of single particles to become correlated. The precise value of $f$ depends on the concentration of mobile ions and the geometry of the host lattice. For this reason, lattice geometry can be key to understanding different behaviours between structural families of solid electrolytes. Explicit interactions between the mobile ions, or between the ions and the lattice, produce additional single-particle correlation effects. These interactions can also promote ordering of the mobile ions, which causes hops by different particles to become correlated. In real solid electrolytes, therefore, microscopic ionic transport depends on both lattice geometry and the nature of interactions acting on the mobile ions.

In this study, we have explored this behaviour for the garnet lattice, which provides a model for the lithium diffusion network in lithium-garnet solid electrolytes. From a theoretical perspective, this lattice possesses intriguing topological features. Previous theoretical and computational analyses of correlated ionic transport in crystalline lattices have considered only lattices where all sites are geometrically equivalent. The garnet lattice, however, contains both four-coordinate tetrahedral sites and two-coordinate octahedral sites, arranged in an open three-dimensional network of interconnected rings (cf.\ Fig.~\ref{fig:garnet_network_schematic}). 

To study correlation effects in the garnet lattice, we have performed  lattice-gas kinetic Monte Carlo simulations.\cite{Morgan_JOSS2017} These consider the host structure as an idealised lattice, and describe ion interactions through simple model Hamiltonians. Rather than seeking an explicit description of a single material, as one might by using e.g.\  first-principles or classical molecular dynamics,\cite{JalemEtAl_ChemMater2013,MeierEtAl_JPhysChemC2014,BurbanoEtAl_PhysRevLett2016} here we focus on understanding general behaviour as a function of lattice geometry and mobile ion stoichiometry, and how this changes in response to conceptually simple, but physically motivated, microscopic interactions.

We find that for the non-interacting (volume exclusion only) system, the single particle correlation effects due to the lattice geometry are more significant than for any previously studied three-dimensional lattice (Table~\ref{tab:vacancy_correlation_factors}). We propose that this is a consequence of the lattice containing two-coordinate octahedral sites, which act as bottlenecks to diffusion, producing correlation effects intermediate between those of simple three-dimensional and one-dimensional lattices. 

Explicit interactions acting on the mobile ions (here, nearest-neighbour repulsion and site-occupation energy differences) produce stronger single-particle correlation effects with a complex variation with $\xLi$. These explicit interactions also promote ordering at set mobile ion concentrations, which manifests as large collective correlation effects, with $f_\mathrm{I}\to0$ as $T\to0$.\cite{Murch_SolStatIonics1982} The precise mobile-ion stoichiometry where ordering occurs depends on the lattice geometry and the explicit form of the interaction energy term. Ordering occurs for mobile-ion stoichiometries that are commensurate with the stoichiometry and symmetry of lattice sites, where ordering minimises the ion-interaction energy. In the cases considered here, nearest-neighbour repulsion promotes ordering at $\xLi=6$, with all octahedral sites occupied and all tetrahedral sites vacant. A site-occupation energy that favours tetrahedral site occupation promotes ordering at $\xLi=3$, with all tetrahedral sites occupied and all octahedral sites vacant. 

The ionic conductivity depends on the mobile ion concentration directly, through the mole fractions of mobile ions and vacant sites, and indirectly, through the collective correlation factor (cf.\ Eqn.~\ref{eqn:sigma_conc_dependence}). Because the form of $f_\mathrm{I}(\xLi)$ depends on the interaction energy term, for interacting systems there is no simple expression that gives the mobile-ion concentration that maximises the ionic conductivity. The simulations presented here show that $\arg \max \sigma^\prime(\xLi)$ is in fact very sensitive to the type and strength of mobile ion interactions, with the ``optimal'' lithium stoichiometry varying from $\xLi=1.5$--$7.5$ within the range of parameters we have considered. Even within the simplified models studied here, therefore, ionic transport on the garnet lattice exhibits correlation effects that are both more significant than predicted for simple three-dimensional lattices, and that show a complex dependence on mobile ion stoichiometry.

\begin{figure}[tb]
  \centering
  \resizebox{8cm}{!}{\includegraphics*{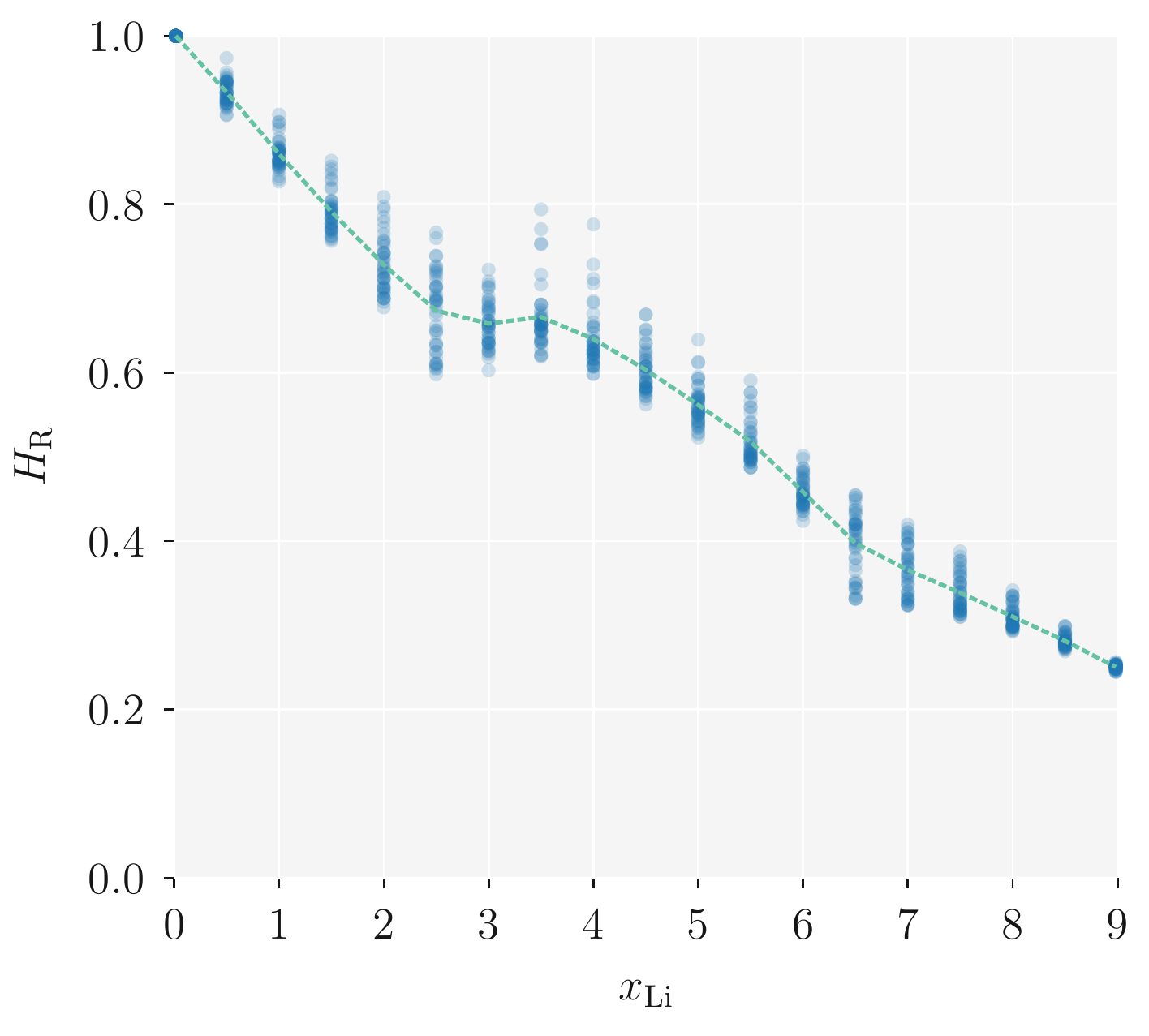}} %
    \caption{\label{fig:haven_ratios}Calculated Haven ratios for all interaction parameter sets considered, as a function of lithium stoichiometry, $\xLi$. The dashed line shows the mean values across these parameter sets.}
\end{figure}

The prediction that lithium garnet solid electrolytes exhibit strong correlation effects is consistent with the observation of highly cooperative diffusion processes in first-principles simulations.\cite{JalemEtAl_ChemMater2013, MeierEtAl_JPhysChemC2014} Because the quantitative correlation behaviour is sensitive to the mobile-ion interactions, this raises the question of how the interactions in real lithium garnet electrolytes might map to the effective interactions considered here. This sensitivity also raises the possibility of tuning ionic conductivities through isovalent substitution within the host lattice. As an example, host lattices containing different  cations will have different lattice parameters, and different distances between neighbouring tetrahedral and octahedral sites. This will modify both $\Delta E_\mathrm{site}$ and $E_\mathrm{nn}$, with consequential non-trivial effects on $f_\mathrm{I}$, and hence on $\sigma$. 

Although $f$ and $f_\mathrm{I}$ are sensitive to the interaction parameters, their ratio, $\frac{f}{f_\mathrm{I}}=H_\mathrm{R}$, is less so. The calculated Haven ratios can therefore be used to improve the quantitative nature of conversions between tracer diffusion coefficients and ionic conductivities, via the modified Nernst-Einstein relation (Eqn.~\ref{eqn:correlated_nernst_einstein}). Fig.~\ref{fig:haven_ratios} shows the calculated Haven ratios for all parameter sets considered in our study. Also plotted is the average Haven ratio across parameter sets as a function of lithium stoichiometry.

\begin{figure}[tb]
  \centering
  \resizebox{8cm}{!}{\includegraphics*{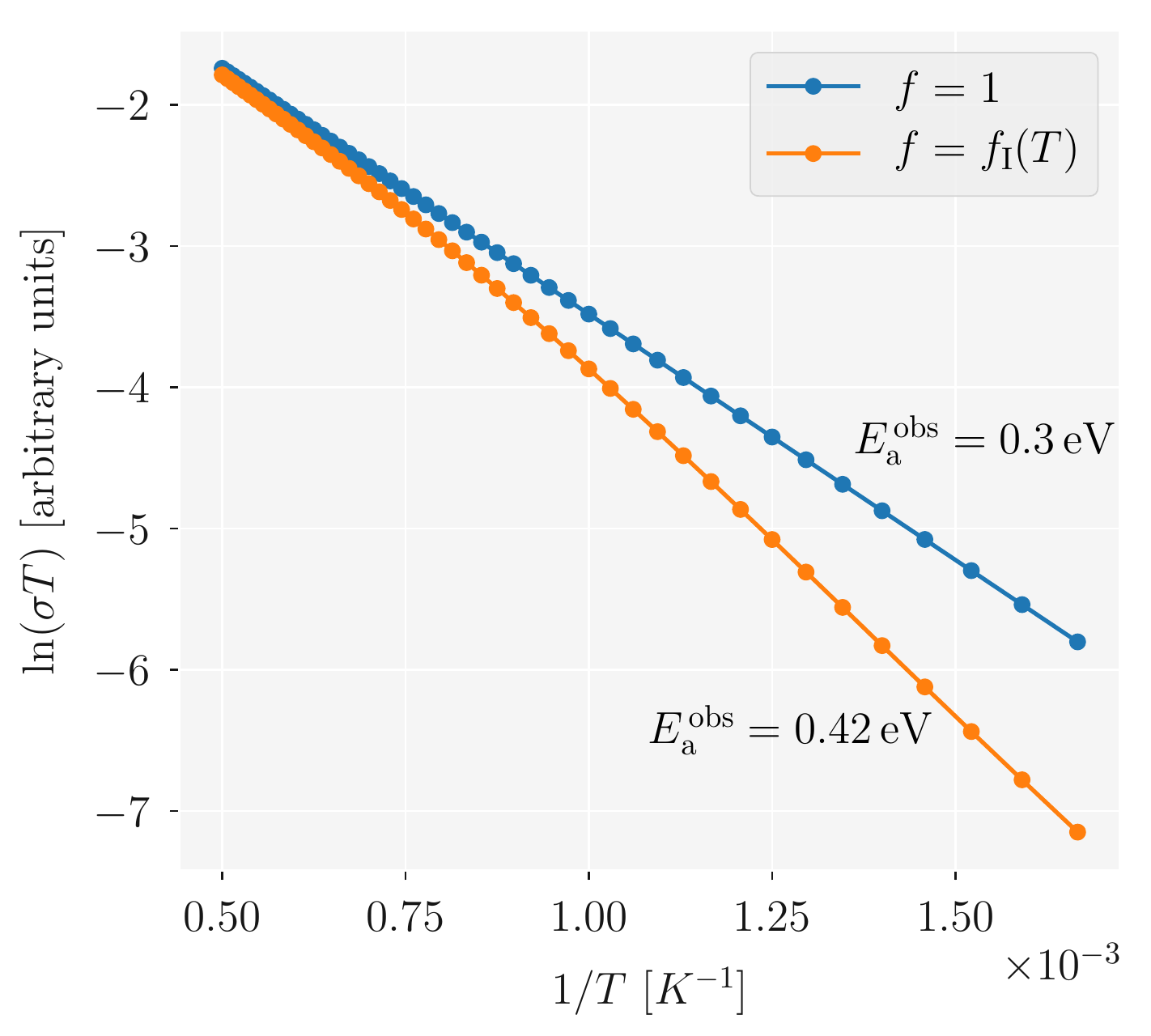}} %
    \caption{\label{fig:arrhenius_example}Relative ionic conductivities calculated for non-interacting particles, using Eqn.~\ref{eqn:random_walk_conductivity}, and for particles subject to nearest-neighbour repulsion, using Eqn.~\ref{eqn:correlated_conductivity}. For the interacting case, $f_\mathrm{I}$ is interpolated from the simulation data described above, with the nearest-neighbour repulsion obtained as the Coulomb energy for two point charges occupying neighbouring sites, using a typical garnet relative permittivity of $\epsilon_\mathrm{r}=50$.\cite{RettenwanderEtAl_InorgChem2015} In each case an ``observed'' activation energy, $E_\mathrm{a}^\mathrm{obs}$, is derived by fitting a straight line to the low temperature $T\leq1000\,\mathrm{K}$ data. Full details of this analysis are available in the supporting dataset.\cite{morgan_benjamin_j_2017_821864}}
\end{figure}

The sensitivity of $f$ and $f_\mathrm{I}$ to the interaction details means that estimating these correlation factors for specific materials purely from this study is challenging. It is clear, however, that assuming that $f=f_\mathrm{I}=1$ is likely to introduce quantitative errors when extrapolating between hop rates and tracer diffusion coefficients or ionic conductivities. One qualitative observation is that where ion interactions are present, the resulting correlation effects will increase as the temperature decreases. One consequence of this temperature-dependent correlation is that Arrhenius plots of tracer diffusion coefficients or ionic conductivities may not give  straight lines. Instead, as temperatures tend to zero, and correlation effects become more significant, they are expected to curve downwards. Fig.~\ref{fig:arrhenius_example} shows Arrhenius plots of relative ionic conductivities at $\xLi=6$, calculated for non-interacting particles, using Eqn.~\ref{eqn:random_walk_conductivity}, and for interacting particles subject to nearest-neighbour repulsion. In both cases a microscopic activation energy of $0.3\,\mathrm{eV}$ is used. For the interacting case, $f_\mathrm{I}$ is interpolated from our simulation results. The nearest-neighbour repulsion energy is chosen as the Coulomb energy for two point charges occupying neighbouring sites, at a separation of $2.4\,\mathrm{\angstrom}$, using a typical garnet relative permittivity of $\epsilon_\mathrm{r}=50$.\cite{RettenwanderEtAl_InorgChem2015} The Arrhenius plot for the data calculated assuming uncorrelated hopping gives a straight line, and a linear fit to obtain an ``observed'' activation energy recovers the microscopic activation energy of $0.3\,\mathrm{eV}$. The data calculated including the temperature-dependent collective correlation factor, however, fall below the first data set---the collective correlation decreases the ionic conductivity relative to the ideal value---and this effect becomes more significant as the  the temperature decreases and the correlation effects strengthen. Fitting to the low temperature regime ($<1000\,\mathrm{K}$) gives an ``observed'' activation energy of $0.42\,\mathrm{eV}$. The additional temperature dependence in the collective correlation means that the observed activation energy can not be directly equated with the microscopic activation energy.\footnote{Equivalent ``super-Arrhenius'' behaviour has been described in the context of diffusion in fragile, supercooled glasses: see e.g.\ Ref.~\onlinecite{SouzaAndWales_PhysRevB2006}.}

One of the limitations of this study is that it uses a fixed predetermined lattice geometry. The ordering predicted at $\xLi=3$ and $\xLi=6$ occurs at  lithium stoichiometries that are commensurate with the lattice symmetry and site ratios. In both cases, the ordered lithium configuration, with either tetrahedral or octahedral sites fully occupied and the alternate site fully vacant, preserves the lattice symmetry. In real materials lattice distortions are possible, and ordering of mobile ions can occur in concert with lattice symmetry breaking. In the lithium garnets the prototypical example is the low-temperature tetragonal phase of \chem{Li_7La_3Zr2O_{12}} (LLZO).\cite{BernsteinEtAl_PhysRevLett2012,AwakaEtAl_ChemLett2011} This material is cubic at high temperature ($T>600\,\mathrm{K}$), but at lower temperatures it undergoes a tetragonal distortion, associated with the lithium ions ordering to occupy all the octahedral sites and one third of the tetrahedral sites, accompanied by a decrease in ionic conductivity of two orders of magnitude. This is another example of ions ordering at low temperature, with low ionic conductivities as a consequence of the resulting strong correlation effects.\cite{BurbanoEtAl_PhysRevLett2016} Again, this ordering occurs at a stoichiometry  commensurate with the lattice symmetry. In the case of LLZO, ordering is promoted at $\xLi=7$ because the accompanying tetragonal distortion lowers the crystal symmetry; in each lattice ring the six tetrahedral sites, equivalent by symmetry in the cubic lattice, become an inequivalent set of (2+4) paired sites. 

Lithium ordering coupled to symmetry breaking has also been predicted at other lithium stoichiometries by Kozinsky \emph{et al.},\cite{KozinskyEtAl_PhysRevLett2016} who performed a group theoretical analysis combined with first-principles energy calculations. Interestingly, this study predicted an ordered phase at $\xLi=6$ with a lower symmetry than the parent cubic phase, with octahedra and tetrahedra occupied in a 3:1 ratio, as well as ordered phases at other lithium stoichiometries, again accompanied by spontaneous symmetry breaking and lattice distortion. Because we have restricted our study to the ideal cubic garnet lattice, our results provide no information about possible alternate ordered phases that might be  energetically favoured in distorted garnet lattices (e.g.\ at $\xLi=6$) or that might appear at stoichiometries where we do not predict ordering. A complete description of the order--disorder phase behaviour in lithium garnets would need to include not only the ideal cubic lattice, but also symmetry-broken lattices. Studying this behaviour within a lattice-gas Monte Carlo simulation scheme would require a more sophisticated approach than used here.

A second limitation is the use of the Metropolis scheme for calculating hopping probabilities (Eqn.~\ref{eqn:metropolis}). This approach considers hops to be barrierless, with hopping probabilities that depend only on the energy differences between initial and final configurations. In real materials, ion hopping is an activated process, and ions move across potential energy barriers. Expressions for hopping probabilities that take these barrier heights into account are expected to give more accurate kinetics, but require parameters that describe typical barrier heights, and how these are affected by the instantaneous local arrangements of mobile atoms.\cite{SingerAndPeschel_ZPhysikBCondensedMatter1980, VanderVenAndCeder_HandbookofMaterialsModelling2010} For specific materials, these parameters can be derived from first-principles calculations.\cite{VenEtAl_PhysRevB2001,VenAndCeder_PhysRevLett2005,LeeEtAl_PhysRevB2011, GrieshammerEtAl_PhysChemChemPhys2014} For this study, we have focussed on a broad description of the geometry effects in the garnet lattice. Including hopping barriers in a general scheme would significantly increase the dimensionality of the available parameter space, making a full analysis of  lattice geometry effects impractical. It is apposite, however, to consider to what extent the results presented here, using the simpler Metropolis scheme, might differ from equivalent calculations that do account for hopping barrier effects. Addressing this question in the context of specific garnet materials will be the subject of a future study. 

Interestingly, the behaviour described here; that in interacting systems, ordering is predicted at particular stoichiometries commensurate with the lattice symmetry, which manifests as strong correlation effects and greatly reduced transport coefficients; is qualitatively similar to results obtained for other lattice geometries using lattice-gas Monte Carlo models that do include barrier terms. For example, Murch and Thorn have modelled the effects of site-energy differences and nearest-neighbour repulsion in the 2D honeycomb lattice, using a fixed transition barrier when deriving their hopping probabilities, and observed ordering and strong correlation effects at half-occupancy.\cite{MurchAndThorn_PhilMag1977a,MurchAndThorn_PhilMag1977b,MurchAndThorn_PhilMag1977c} The observation of qualitatively similar behaviour using models that do account for hopping barriers, albeit for different lattice geometries, suggests that these effects are not strongly dependent on the precise scheme. It should also be noted that any ordering of the mobile particles, which is the physical origin of these correlation effects, is independent of any transition barriers. The equilibrium distribution of particles depends only on the relative energies of different configurations, and is therefore exactly described (for a given Hamiltonian) by the Metropolis scheme. 

A third consideration is that ion transport is assumed to be effected by a sequence of discrete hops made by individual ions. Although this is a good model for ionic transport in a large number of solid electrolytes, this is not always the case. In particular, so-called ``superionic'' solid electrolytes exhibit diffusion mechanisms where ions move through highly concerted ``liquid-like'' processes.\cite{Catlow_AnnRevMaterSci1986,Hull_RepProgPhys2004} For solid lithium-ion electrolytes with particularly high ionic conductivities, such as those typically of interest for  all-solid-state lithium-ion batteries, it is not known to what extent ion transport proceeds by concerted rather than single-ion diffusion mechanisms.\footnote{The highly concerted diffusion mechanisms that characterise superionic electrolytes are also typically associated with significant correlation effects and Haven ratios that deviate from $H_\mathrm{R}=1$.\cite{SalanneEtAl_JPhysChemC2012,Hull_RepProgPhys2004}} In the case of the lithium garnets, data are limited and confined to cubic LLZO. Meier \emph{et al}.\ performed a first-principles metadynamics study of cubic LLZO, and identified a concerted diffusion process in their simulation trajectory,\cite{MeierEtAl_JPhysChemC2014} and a recent first-principles study by He \emph{et al}.\ showed that concerted diffusion processes in this material can have lower potential energy barriers than single-ion hopping processes.\cite{HeEtAl_NatureComm2017} Support for single-ion hopping, however, comes from a study by Chen \emph{et al.}, who performed classical molecular dynamics simulations of LLZO.\cite{ChenEtAl_SciRep2017} By decomposing their simulation trajectories into sequences of single-ion hops, these authors showed that diffusion can be modelled as a Poisson process, which is a characteristic signature of an independent hopping process.\cite{MorganAndMadden_PhysRevLett2014}

The question of contributions from concerted diffusion processes is not only pertinent to high conductivity systems, but can also be important in ordered phases with low ionic conductivities. Under strong ordering of mobile ions, correlation effects may sufficiently impede ion transport by single particle hopping that alternate concerted mechanisms become the dominant ion transport process.\cite{MorganAndMadden_PhysRevLett2014} This is believed to be the case for the low-temperature tetragonal phase of LLZO, with lithium transport effected by highly concerted motion of groups of ions moving around the lattice rings. \cite{BurbanoEtAl_PhysRevLett2016} In the context of developing a theoretical framework that can quantitatively connect microscopic diffusion processes in solid electrolytes to macroscopic transport coefficients, a general treatment of concerted diffusion mechanisms remains an intriguing problem. 

\section{Data Availability}
Supplementary material for this study is available as a GitHub repository, published under the CC-BY-SA-4.0 license.\cite{morgan_benjamin_j_2017_821864} This repository contains (1) the complete data set used to support the findings of this study, (2) example scripts for running \texttt{lattice\_mc} simulations on a garnet lattice and collating output data, and (3) a Jupyter notebook containing the code used to generate Figs.\ \ref{fig:non-interacting_data}--\ref{fig:conductivity_miniplots}. The \texttt{lattice\_mc} code is available under the MIT license.\cite{Morgan_JOSS2017}

 

\emph{Data accessiblity.} All data used to support the findings of this study are available as part of the Supplementary Material dataset available at http://dx.doi.org/10.5281/zenodo.821870.\\
\emph{Competing interests.} B.~J.~M.\ has no competing interests.\\
\emph{Authors' contributions.} Does not apply.\\
\emph{Funding.} B.~J.~M.\ acknowledges support from the Royal Society (UF130329).\\
\emph{Research ethics.} Does not apply.\\
\emph{Animal ethics.} Does not apply.\\
\emph{Permission to carry out fieldwork.} Does not apply.\\
\emph{Acknowledgements} B.~J.~M.\ would like to thank M.~Burbano and M.~Salanne for stimulating discussions related to this work.

\section{Appendix}

\renewcommand{\thefigure}{A\arabic{figure}}
\setcounter{figure}{0}

\begin{figure*}[tb]
  \centering
  \resizebox{16cm}{!}{\includegraphics*{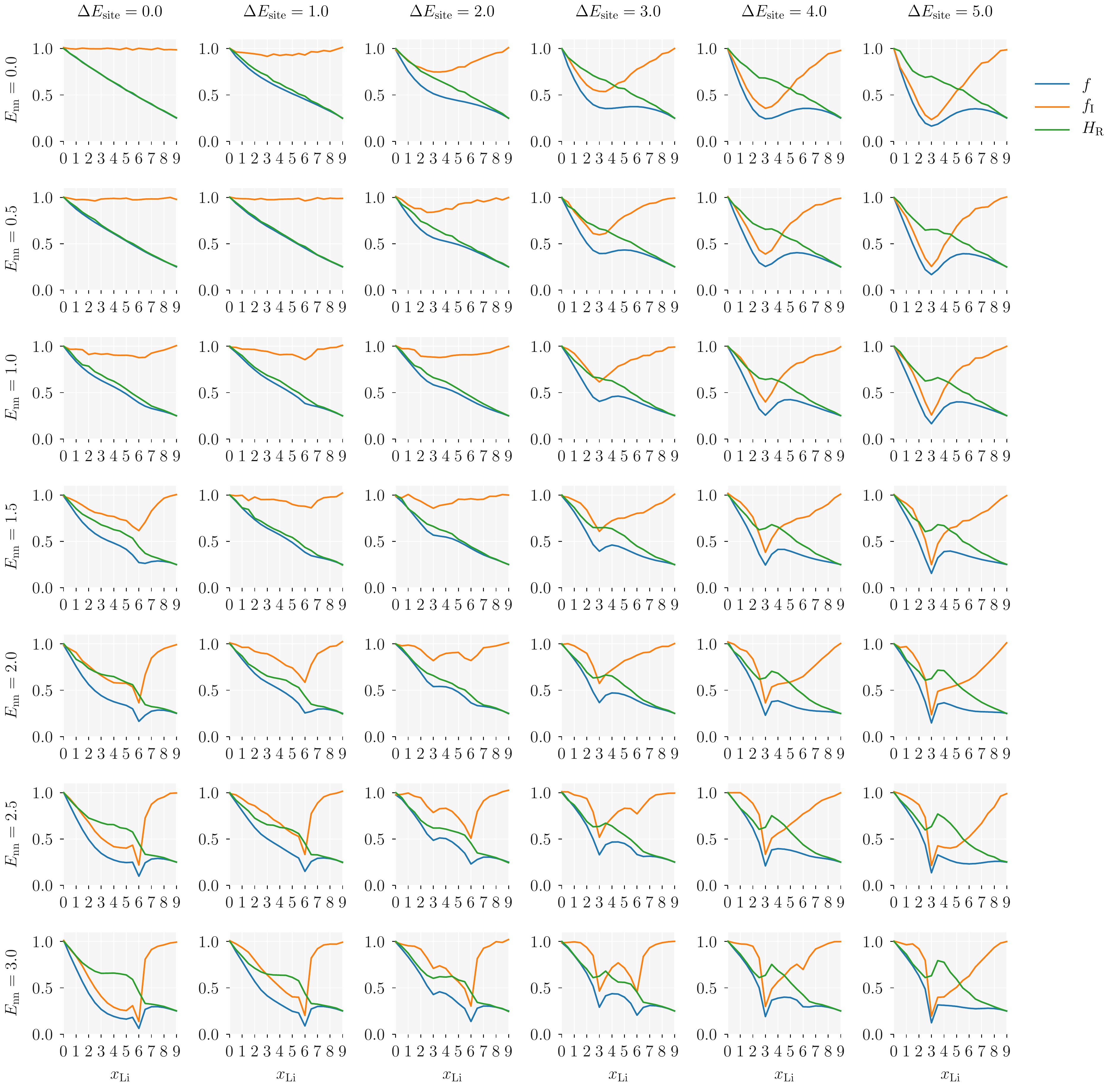}} %
    \caption{\label{fig:correlation_miniplots}Correlation factors for the garnet lattice as a function of mobile-ion stoichiometry, $\xLi$, nearest-neighbour repulsion, $E_\mathrm{nn}$, and site-occupation energy difference, $\Delta E_\mathrm{site}$. Each subplot shows the single particle correlation factor, $f$, the collective correlation factor, $f_\mathrm{I}$, and the Haven ratio, $H_\mathrm{R}$.}
\end{figure*}

\begin{figure*}[tb]
  \centering
  \resizebox{16cm}{!}{\includegraphics*{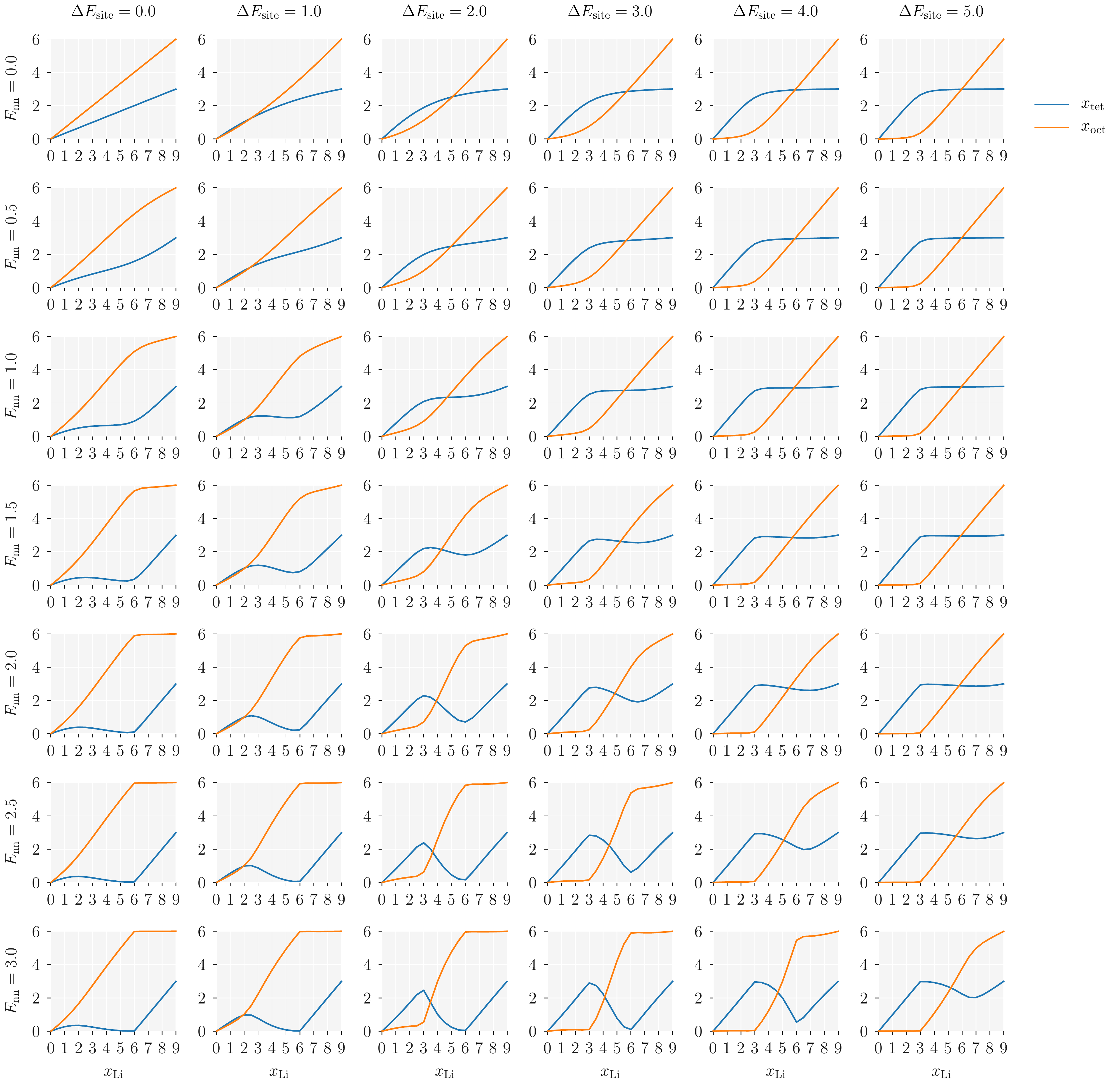}} %
    \caption{\label{fig:site_occupation_miniplots}Average site occupations for the garnet lattice as a function of mobile-ion stoichiometry, $\xLi$, nearest-neighbour repulsion, $E_\mathrm{nn}$, and site-occupation energy difference, $\Delta E_\mathrm{site}$. Each subplot shows the time-averaged number of occupied tetrahedral, $x_\mathrm{tet}$, and octahedral, $x_\mathrm{oct}$, sites per formula unit.}
\end{figure*}

\begin{figure*}[tb]
  \centering
  \resizebox{16cm}{!}{\includegraphics*{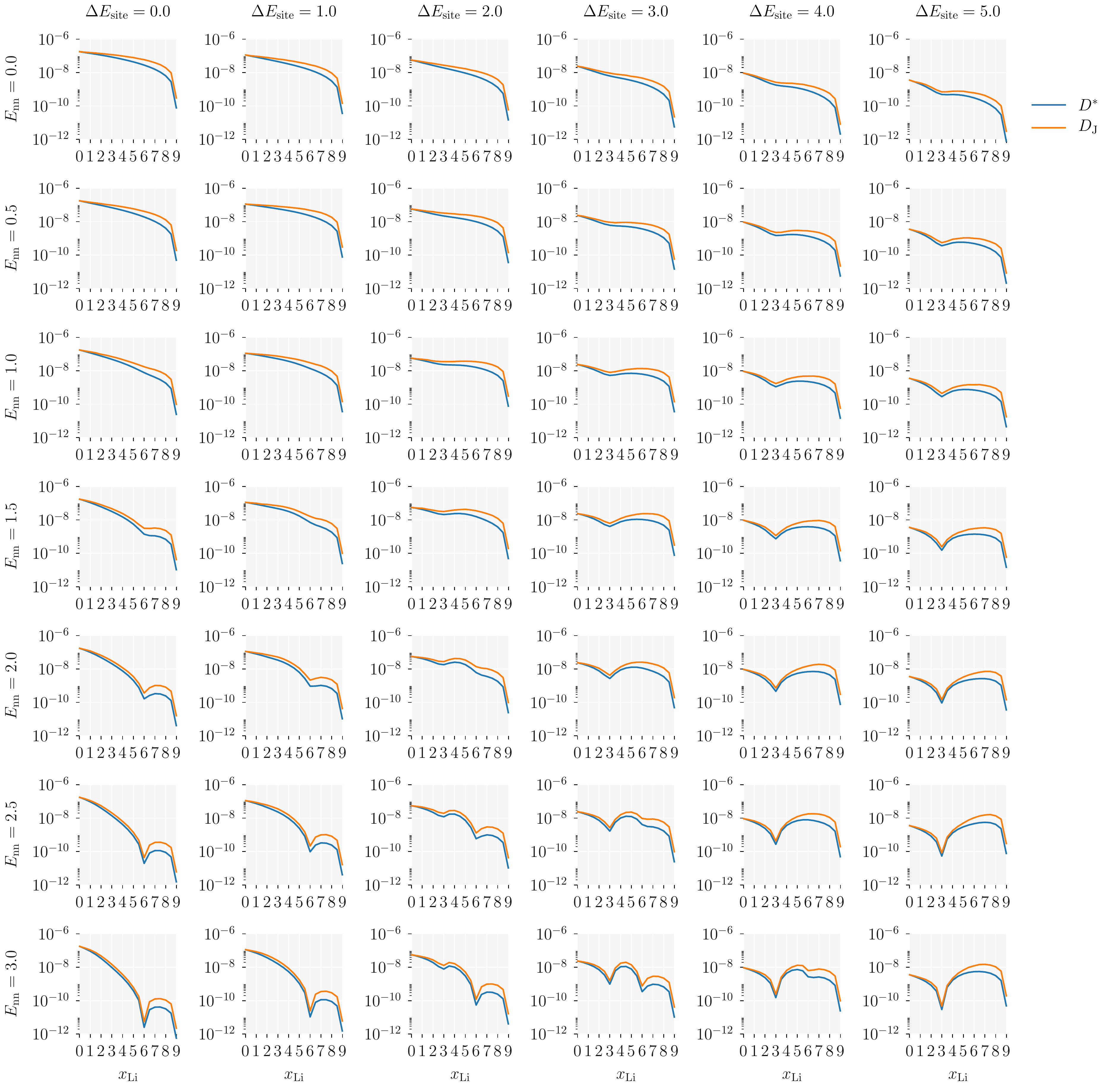}} %
    \caption{\label{fig:diffusion_miniplots}Diffusion coefficients for the garnet lattice as a function of mobile-ion stoichiometry, $\xLi$, nearest-neighbour repulsion, $E_\mathrm{nn}$, and site-occupation energy difference, $\Delta E_\mathrm{site}$. Each subplot shows the tracer diffusion coefficient, $D^*$, and the collective ``jump'' diffusion coefficient, $D_\mathrm{J}$.}
\end{figure*}

\begin{figure*}[tb]
  \centering
  \resizebox{16cm}{!}{\includegraphics*{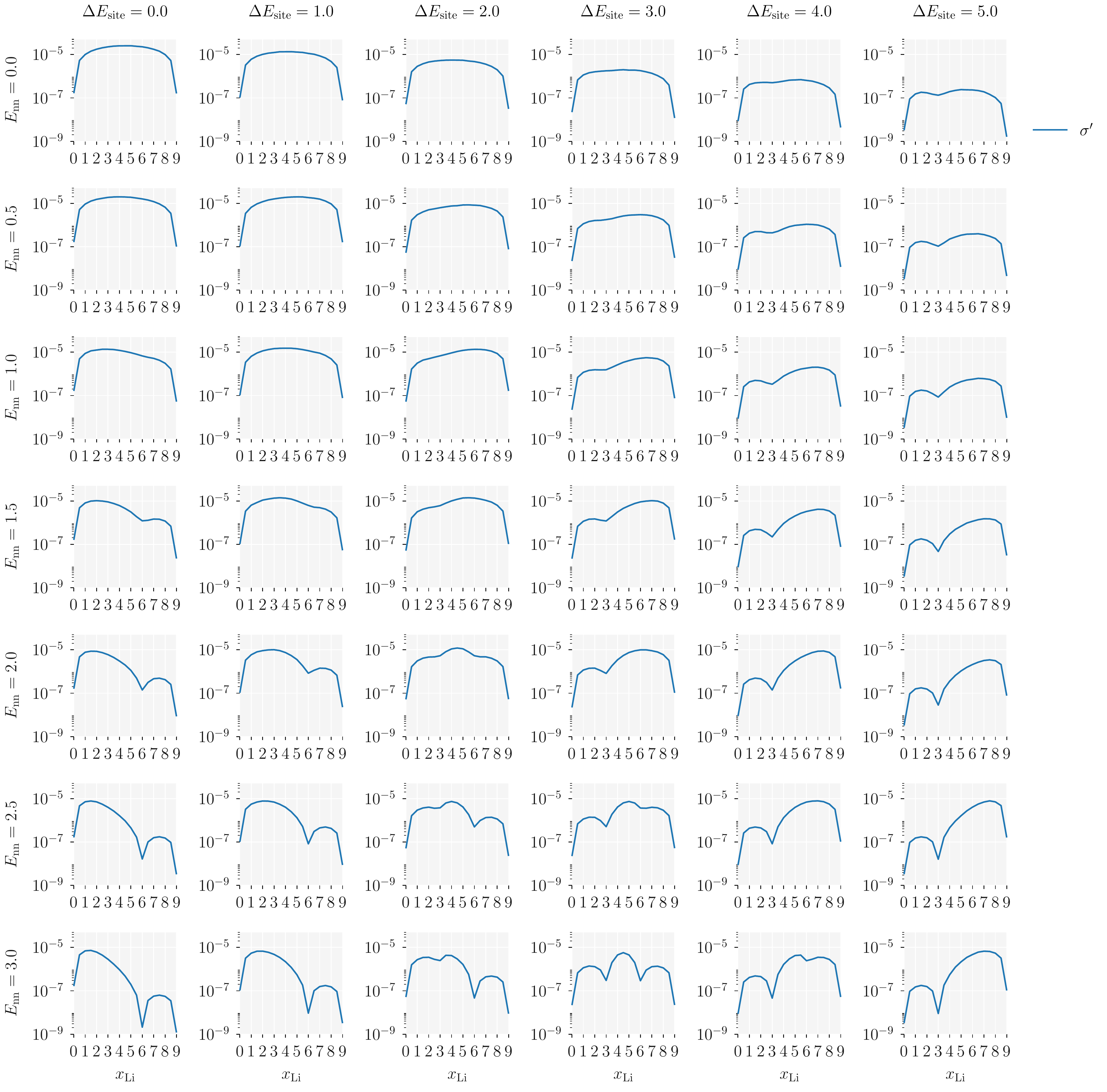}} %
    \caption{\label{fig:conductivity_miniplots}Reduced ionic conductivity for the garnet lattice as a function of mobile-ion stoichiometry, $\xLi$, nearest-neighbour repulsion, $E_\mathrm{nn}$, and site-occupation energy difference, $\Delta E_\mathrm{site}$. Each subplot shows the reduced ionic conductivity, $\sigma^\prime$ (Eqn.~\ref{eqn:reduced_sigma}).}
\end{figure*}


\bibliography{Bibliography}
\end{document}